%% file: dressi_eg2022.tex
\documentclass{egpubl}
\usepackage{eg2022}

\ConferencePaper        
\usepackage[utf8]{inputenc}    
\usepackage[T1]{fontenc}
\usepackage{dfadobe}

\biberVersion
\BibtexOrBiblatex
\usepackage[backend=biber,bibstyle=EG,citestyle=alphabetic,backref=true]{biblatex}
\electronicVersion
\PrintedOrElectronic

\ifpdf \usepackage[pdftex]{graphicx} \pdfcompresslevel=9
\else \usepackage[dvips]{graphicx} \fi

\usepackage{egweblnk}

\usepackage{amsmath}
\usepackage{amsfonts}
\usepackage{subfiles}
\usepackage{multirow}
\usepackage{comment}
\usepackage{listings}
\usepackage{here}
\usepackage{algorithm}
\usepackage{algorithmicx}
\usepackage{algpseudocode}
\usepackage{svg}
\usepackage{subcaption}
\usepackage{makecell}
\usepackage{booktabs}
\usepackage{pifont}
\usepackage{graphicx}
\usepackage{color}
\usepackage{colortbl}
\usepackage{arydshln}
\usepackage{textcomp}

\DeclareMathOperator*{\argmin}{arg\,min}

\newcommand{\figurename}{Fig.}
\newcommand{\secname}{Section}
\newcommand{\subsecname}{Section}
\newcommand{\subsubsecname}{Section}

\newcommand{\BeginFigure}{\begin{figure}[tb]}
\newcommand{\EndFigure}{\end{figure}}

\newcommand{\BeginFigureTwoCol}{\begin{figure*}[tb]}
\newcommand{\EndFigureTwoCol}{\end{figure*}}

\newcommand\acksname{Acknowledgments}
\specialcomment{acks}{%
  \begingroup
  \section*{\acksname}
  \phantomsection\addcontentsline{toc}{section}{\acksname}
}{%
  \endgroup
}

\newcommand{\diffred}[1]{#1}

\title[Dressi]
{Dressi: A Hardware-Agnostic Differentiable Renderer with Reactive Shader Packing and Soft Rasterization}

\input{./misc/author_eg}

\begin{document}

\teaser{
  \includegraphics[width=\linewidth]{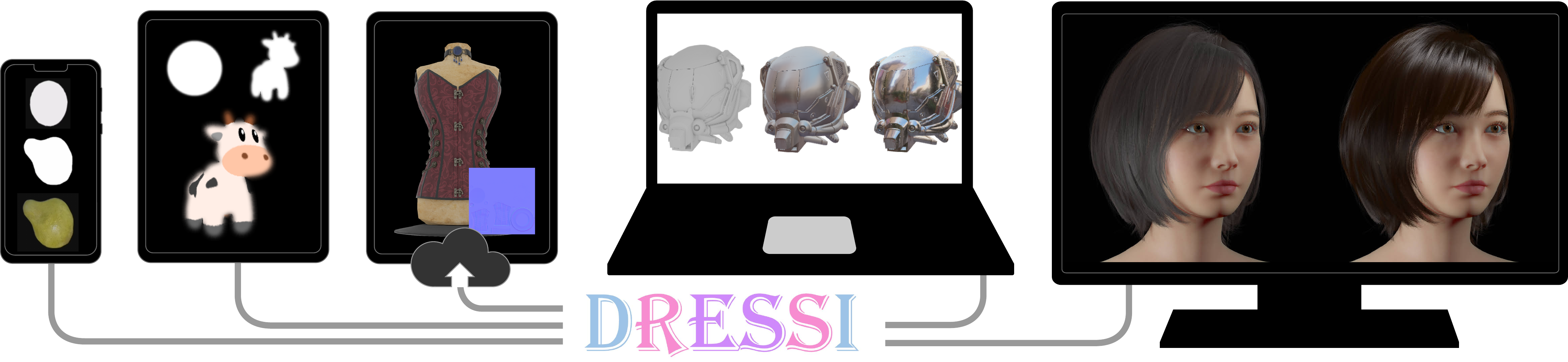}
  \centering
  \caption{Applications of our differentiable renderer called {\textit{Dressi}}. From left to right: optimization of geometry and material (a pear and a cow), normal map, environment map and material, and hair material of the digital human. The applications work on all devices supporting a modern graphics API Vulkan \cite{vulkan} such as mobiles, tablets, cloud servers, laptops, and desktop PCs.}
  \label{fig:teaser}
}

\maketitle

\input{./0_abstract/abstract}

\def\thefootnote{*}\footnotetext{These authors contributed equally to this work}


\renewcommand{\ttdefault}{cmtt}
\normalfont

\input{1_introduction/introduction}
\input{2_related_work/related_work}

\input{3_dressi/dressi}

\input{4_validation/validation}
\input{5_experiments/experiments}

\input{6_applications/applications}

\input{7_conclusion/conclusion}
\input{misc/acknowledgement}



\printbibliography

\clearpage
\appendix
\input{9_appendix/appendix}


\end{document}

%% file: misc/author_eg.tex
\author[Y. Takimoto, H. Sato, H. Takehara, K. Uragaki, T. Tawara, X. Liang, K. Oku, W. Kishimoto \& B. Zheng]
{\parbox{\textwidth}
{\centering
  Yusuke Takimoto$^*$,\hspace{1pt}
  Hiroyuki Sato$^*$,\hspace{1pt}
  Hikari Takehara$^*$,\hspace{1pt}
  Keishiro Uragaki,\hspace{1pt}
  \\
  Takehiro Tawara,\hspace{1pt}
  Xiao Liang,\hspace{1pt}
  Kentaro Oku,\hspace{1pt}
  Wataru Kishimoto,\hspace{1pt}
  and Bo Zheng
}\footnotetext{* Equal contribution}
\\
{\parbox{\textwidth}
{\centering
  Huawei Technologies Japan K.K.
}
}
}

%% file: 0_abstract/abstract.tex
\begin{abstract}
%
Differentiable rendering (DR) enables various computer graphics and computer vision applications through gradient-based optimization with derivatives of the rendering equation.
Most rasterization-based approaches are built on general-purpose automatic differentiation (AD) libraries and DR-specific modules handcrafted using CUDA.
Such a system design mixes DR algorithm implementation and algorithm building blocks, resulting in hardware dependency and limited performance.
%
In this paper, we present a practical hardware-agnostic differentiable renderer called {\textit{Dressi}}, which is based on a new full AD design.
The DR algorithms of Dressi are fully written in our Vulkan-based AD for DR, {\textit{Dressi-AD}}, which supports all primitive operations for DR.
Dressi-AD and our {\textit{inverse UV}} technique inside it bring hardware independence and acceleration by graphics hardware.
    {\textit{Stage packing}}, our runtime optimization technique, can adapt hardware constraints and efficiently execute complex computational graphs of DR with reactive cache considering the render pass hierarchy of Vulkan.
    {\textit{HardSoftRas}}, our novel rendering process, is designed for inverse rendering with a graphics pipeline.
Under the limited functionalities of the graphics pipeline, HardSoftRas can propagate the gradients of pixels from the screen space to far-range triangle attributes.
%
Our experiments and applications demonstrate that Dressi establishes hardware independence, high-quality and robust optimization with fast speed, and photorealistic rendering.
%
%

\ccsdesc[500]{Computing methodologies~Rasterization}
\ccsdesc[500]{Computing methodologies~Shape inference}

\printccsdesc

\end{abstract}


%% file: 1_introduction/introduction.tex
\section{Introduction}
\label{sec:introduction}

%
Inverse rendering is a long-standing problem in estimating scene attributes from 2D images in computer graphics and computer vision fields.
Differentiable rendering (DR) plays an important role in applications, such as facial geometry reconstruction \cite{Egger20203DMF, gecer2019ganfit, gecer2021fastganfit} and camera pose estimation \cite{Rhodin:2015}.
It recovers scene parameters by the gradient propagation of the loss functions defined between the rendered and observed images.
For example, it is possible to estimate materials if the appearance and other scene settings are already known \cite{Azinovic2019InversePathTracing}.

%
DR can be grouped into two categories: ray tracing \cite{Li:2018:DMC, NimierDavidVicini2019Mitsuba2}
and rasterization \cite{kato2018renderer, chen2019dibrender, TensorflowGraphicsIO2019}.
The ray tracing-based methods consider complex effects generated by light transport simulations, such as indirect illumination and polarization \cite{NimierDavidVicini2019Mitsuba2}.
They achieve the best rendering quality and accuracy with heavy computation.
In contrast, rasterization-based methods require fewer computations and provide more efficient solutions, attracting much attention for practical scene reconstruction.
We focus on the rasterization-based methods in this paper.

%

%
Most existing rasterization-based DR systems ignore hardware dependencies.
They are designed to run on high-end hardware from specific vendors (e.g., NVIDIA), and they often implement the DR algorithm using automatic differentiation (AD) libraries for neural networks (e.g., PyTorch and TensorFlow) and additional modules written in the general-purpose graphical processing unit (GPGPU) API, which relies on CUDA for most cases.
Because some DR-specific and performance-critical functions (e.g., rasterization and texture sampling) involve pixel-wise operations, writing them down in the DR layer using an efficient batch operation with AD libraries is difficult.
Such a tightly coupled DR system design increases hardware dependency and limits performance.
The AD libraries are not optimized for complex computational graphs inherent to DR.
Moreover, DR systems cannot optimize performance at the border between the AD libraries and handwritten CUDA modules.
Therefore, it is not easy to run the existing DR systems on graphics hardware from various vendors (e.g., Intel, AMD, and Arm) at practical speeds, especially on low-end edge devices.
Communicating with a remote server that has a DR system installed is an option; nonetheless, it impairs the interactivity of real-time applications, owing to latency and raises privacy issues.
Making a DR work efficiently on low-end hardware is a challenging problem.

%
To obtain hardware independence for DR systems, we can use the graphics pipeline API, which has hardware rasterizers and shaders, instead of the GPGPU API.
However, propagating gradients in a screen space under the limitations of graphics pipelines is another challenging problem.
The existing rasterization-based DR methods generate gradients in the screen spaces using the special rendering processes easily implemented by CUDA, which allows developer descriptions with high degrees of freedom; however, their efficient implementation using a graphics pipeline is difficult.
For instance, SoftRasterizer (SoftRas) \cite{liu2019softras} computes distances between pixels and the edges of projected triangles in a pixel-wise manner.
Nevertheless, pixel-wise computation considering all triangles is not easy for hardware rasterizers.
Nvdiffrast \cite{Laine2020diffrast} employs analytic anti-aliasing (AA), which considers edge information among geometrically neighboring triangles.
Unfortunately, accessing mesh topology information is also difficult for shaders.
Therefore, we cannot port the existing methods to the graphics pipeline.

%
We propose Dressi, a practical rasterization-based differentiable renderer that solves the above problems.
Dressi is based on a new design that completely separates the AD layer and DR algorithms.
DR algorithms are fully described by the AD layer and receive hardware independence and performance optimization through the overall system.
Our AD layer is Dressi-AD, which is implemented in Vulkan \cite{vulkan}, a vendor-independent graphics pipeline API.
Furthermore, it is tailored to the DR to support all its primitive operations, including rasterization and texture sampling.
The inverse UV technique in Dressi-AD carefully handles the backward pass of hardware texture sampling.
At runtime, our stage packing dynamically converts the computational graphs into optimal execution orders to the command buffers.
It maximizes hardware performance by considering the constraints for the render pass and subpass of the running hardware.
Static values on computational graphs are often observed in DR, such as rasterization results with static geometries.
Our reactive cache embedded in the stage packing automatically detects the static values and caches them to skip later computations.

%
Furthermore, we propose a new rasterization-based DR method, HardSoftRas, to realize far-range gradients from the screen space to vertex attributes, such as SoftRas, under the limitations of a graphics pipeline.
\diffred{In this paper, we use the term, \textit{far-range} gradient, if the DR system propagates the gradient at a rendered pixel to vertices projected onto far positions in screen space (i.e., more than one pixel away).
  Otherwise, we call it \textit{near-range} gradient.}
The key to the gradient backward in the screen space is the pixel-to-triangle distance computation; however, no prior art has implemented it on a graphics pipeline owing to the difficulty in handling the association between pixels and triangles.
Our face-wise approach, implemented on an inflexible graphics pipeline, successfully updates the pixel-wise computation of the existing methods written in flexible CUDA.
Furthermore, a new depth-shift method is proposed to fit a few buffers generated by the hardware rasterizer, whereas the existing methods rasterize a large number of buffers using software.
We also show that HardSoftRas is a natural extension of anti-aliasing, which is often used in real-time forward rendering.

%
We experimentally show that Dressi behaves identically on various hardware shipped from different vendors (e.g., NVIDIA, AMD, Intel, and Arm) with varying performance, ranging from servers to mobile platforms.
We also validate that our stage packing and reactive cache increase forward and backward speeds.
Moreover, we compare our method to state-of-the-art rasterization-based differentiable renderers to validate HardSoftRas and the overall system.
Dressi achieves better speed and quality with both synthetic and real data.
Finally, we show that our method can render a photorealistic digital human and optimize hair parameters, which are not supported by existing DR systems.

%% file: 2_related_work/related_work.tex
\section{Related Work}
\label{sec:related_work}

\subsection{Ray Tracing-Based Differentiable Rendering}
\label{subsec:ray_dr}
%
Ray tracing-based DR can simulate complex light transport models \diffred{with visibility handling.}
Redner \cite{Li:2018:DMC} established a method for complex inverse problems of scene parameters based on Monte Carlo ray tracing \cite{Kajiya:1986}.
Mitsuba 2 \cite{NimierDavidVicini2019Mitsuba2} can simulate the complicated light transport phenomena such as scattering, polarization, and spectroscopy.
Moreover, it achieves highly efficient computations with template meta-programming for various data types and a retargetable just-in-time (JIT) compilation for AD.
Radiative backpropagation \cite{NimierDavid2020RadiativeBackpropagation} is a memory-efficient adjoint approach that reduces the computation of backward functions in continuous light transportation.
\diffred{Path-space DR \cite{Zhang:2020:PSDR} shows better efficiency using the new Monte Carlo estimators.}
Path replay backpropagation ~\cite{Vicini2021PathReplay} proposes a ray tracing-based mega-kernel generation that maintains linear time complexity with constant memory usage.
Ray tracing-based methods focus on photorealistic results and accurate backpropagation.
Although they have shown remarkable speedups recently, they still require exceedingly high computational costs for low-end devices \diffred{lacking hardware ray tracers.}
Some code optimization techniques that reduce heavy computations on high-end hardware \cite{NimierDavidVicini2019Mitsuba2, Vicini2021PathReplay} are related to our stage packing process.
However, ours differs from existing techniques because it is designed to adapt to varying graphics hardware constraints.

\subsection{Rasterization-Based Differentiable Rendering}
\label{subsec:rast_dr}
%
Rasterization-based DR ~\cite{Loper:ECCV:2014, kato2018renderer, chen2019dibrender, TensorflowGraphicsIO2019} exhibits a faster computational speed than the ray tracing-based DR.
SoftRas ~\cite{liu2019softras} proposes shape optimization as a probabilistic process.
It computes the minimum pixel-edge distance for all pairs of edges and pixels in the screen space.
The distances and depth values are stored in multiple buffers.
Then, the distances are converted to probability, and pixel colors are calculated by aggregating the buffers based on the probability and the depth values.
Although SoftRas can propagate screen space gradients to far-range vertex attributes, it is difficult to handle large geometries because it requires as many buffers as faces.
To improve the scalability against large geometries, PyTorch3D \cite{ravi2020pytorch3d} extends SoftRas by introducing thresholds for the radius of blurring in the screen space and the number of buffers per pixel.
During rasterization, PyTorch3D enlarges each face in the screen space within the blur radius from their edges and stores the distances and depth values per pixel in the buffers.
The pixel-edge distance computation step of SoftRas and PyTorch3D is implemented as a pixel-wise computation to track the triangle correspondences in CUDA.
We cannot port the algorithm to a graphics pipeline because the hardware rasterizer is based on a face-wise calculation.
Another option is to use compute shaders; nonetheless, their computational costs are prohibitive, \diffred{and they are not supported well on mobile platforms.}

%
Nvdiffrast ~\cite{Laine2020diffrast} achieves significant acceleration, identifying modular primitives of DR.
It leverages the performance of GPUs, especially hardware rasterization using OpenGL.
Nvdiffrast generates screen space gradients via analytic anti-aliasing.
In contrast to SoftRas and PyTorch3D, forward rendering of Nvdiffrast retains the original appearance of rendered images.
However, its screen space gradients are propagated only within near-range vertices, because anti-aliasing focuses on boundaries.
Moreover, its anti-aliasing step handles edge-triangle correspondences in the CUDA code.
Operating edge-based data structures is difficult for the shaders of graphics pipelines.
Nvdiffmodeling \cite{Hasselgren2021} is an optimization method for a mesh and shading model with physically-based shading over Nvdiffrast.
Nvdiffmodeling reconstructs transparency using depth peeling \cite{everitt2001interactive}, but depth peeling is not used for mixing far-range triangle attributes, as in our study.
A survey paper \cite{Kato2020DifferentiableRenderingSurvey} summarizes recent DR methods.

\subsection{Automatic Differentiation}
\label{subsec:autodiff}
%
AD automatically calculates function derivatives.
Advances in the usability of AD ~\cite{Griewank2008EvaluatingDerivatives, ceres-solver} gives rise to new concepts of differential programming, such as DR.
AD on GPU utilizes implementation techniques \diffred{such as shader code generation ~\cite{lattner2008llvm, syoyo2008Mudalang, guenter2011symbolic, muranushi2012paraiso, Devito2017Opt, he2018slang}} and kernel fusion ~\cite{Wang2010KernelFusionEffective, cupy_learningsys2017}.
Checkpointing ~\cite{Griewank1992AchievingLogarithmicGrowth} is a method that reduces memory usage in a reverse-mode AD.
%
Although recent deep learning compilers utilize compilation techniques ~\cite{li2021deepLearningCompiler}, they do not sufficiently support general differentiable programming.
Most deep networks are dense and stationary because they consist of block-shaped modules.
In contrast, computational graphs in other fields, such as DR and inverse physics simulation ~\cite{hu2019difftaichi}, can be complicated, sparse, and dynamic.
Most rasterization-based DRs develop DR algorithms on AD libraries for deep learning, such as PyTorch \cite{NEURIPS2019_9015} and TensorFlow \cite{tensorflow2015-whitepaper}, and additionally implement manual CUDA modules for DR-specific rasterization and texture sampling.
Therefore, their performance is not well optimized.
Unique JIT compilation methods are proposed in each domain.
Mitsuba 2 ~\cite{NimierDavidVicini2019Mitsuba2} has proposed the computational graph simplification for DR that repeats kernel fusion before JIT compilation.
AsyncTaichi ~\cite{hu2020AsyncTaichiWholeProgramOptimizations} proposes asynchronous front-end JIT optimization for dynamic optimization tasks.
%
From the viewpoint of supporting platforms, standard AD libraries require GPGPU, particularly CUDA, and they can run on a CPU.
This indicates that they do not consider desktop PCs and laptops with other vendors' GPUs and mobiles.
Mitsuba 2 achieves stronger hardware independence, thanks to its own JIT compilation.
It works either on a GPU with CUDA or various CPUs with single instruction/multiple data (SIMD) operations.
However, they do not consider other GPU platforms.
We perform a JIT compilation of AD for DR applications that works on all modern GPUs.
TensorFlow.js \cite{smilkov2019tensorflow} has an OpenGL Shading Language (GLSL) backend for cross-platform browsers, but its performance for DR is limited because it is designed for neural networks.
\diffred{Enzyme \cite{NEURIPS2020_9332c513} supports AD of low-level virtual machine (LLVM) codes.
  Enzyme with a translator between LLVM and standard portable intermediate representation (SPIR)-V would be an option for hardware-agnostic DR, but SPIR-V does not contain a hardware rasterizer.
  Thus, it is difficult for Enzyme to establish a high-performance DR.}

%
\BeginFigureTwoCol
\centering
\includegraphics[width=\textwidth]{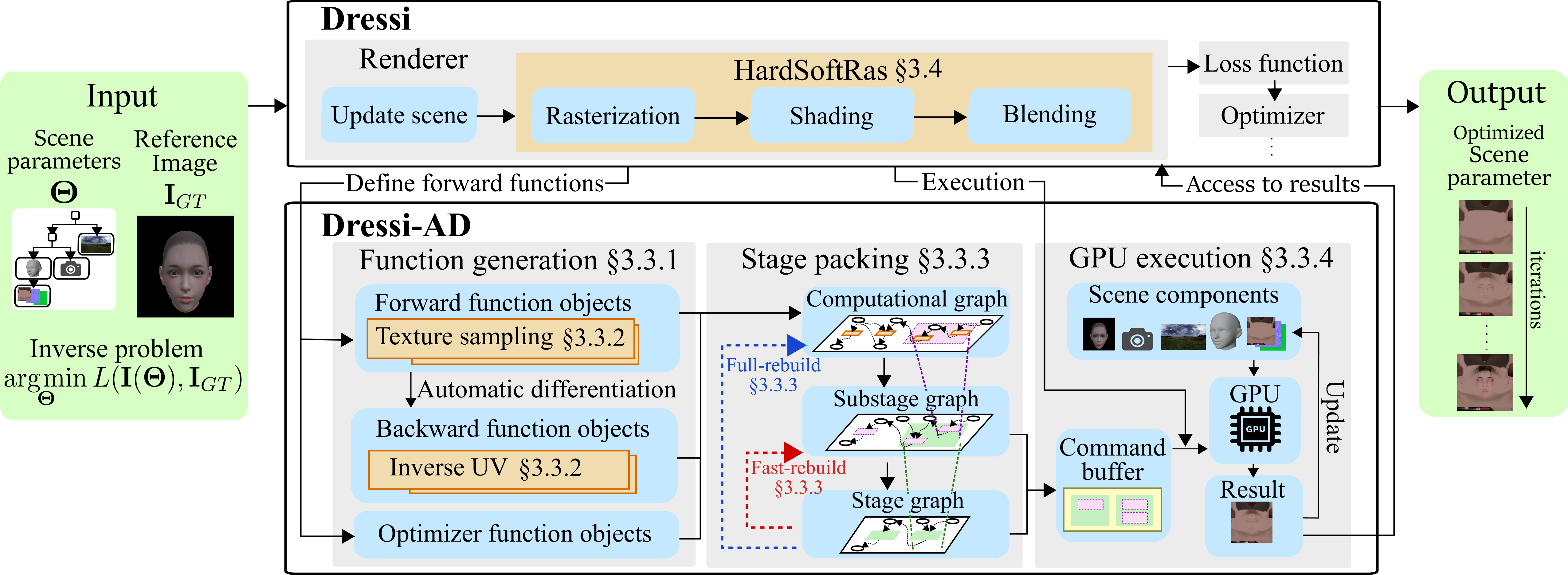}
\caption{Dressi's pipeline.
  This figure shows a toy inverse problem for the optimization of the albedo texture in a set of scene parameters, ${\Theta}$, to fit the rendered result to the reference image, ${{\mathbf{I}}_{GT}}$.
  Dressi is written with forward functions of Dressi-AD.
  Dressi-AD builds a computational graph and executes inverse rendering on the GPU via function generation and stage packing.
}
\label{fig:pipeline}
\EndFigureTwoCol
%
%

%
\input{supplements/system_comparison/table}

%% file: supplements/system_comparison/table.tex
\begin{table*}[t]
    \footnotesize
    \center
    \begin{tabular}{l|cccc}
                          & SoftRas ~\cite{liu2019softras}                   & PyTorch3D ~\cite{ravi2020pytorch3d} & Nvdiffrast ~\cite{Laine2020diffrast} & \textbf{Ours}                \\ \hline
        Graphics Hardware & NVIDIA                                           & NVIDIA                              & NVIDIA                               & NVIDIA, AMD, Intel, Arm, ... \\ \hline
        AD                & PyTorch                                          & PyTorch                             & PyTorch and TensorFlow               & Dressi-AD                    \\ \hline \hline
                          & \multicolumn{4}{c}{Implementation / Backend API}                                                                                                             \\ \hline \hline
        Math              & AD / CUDA                                        & AD / CUDA                           & AD / CUDA                            & AD / Vulkan                  \\
        Rasterization     & Manual / CUDA                                    & Manual / CUDA                       & Manual / CUDA + OpenGL               & AD / Vulkan                  \\
        Texture sampling  & Manual / CUDA                                    & AD / CUDA                           & Manual / CUDA                        & AD / Vulkan                  \\ \hline
    \end{tabular}
    \caption{\label{tab:system_comparison} Comparison of rasterization-based DR frameworks.
        Thanks to our full AD approach with Dressi-AD based on Vulkan, ours can work on various graphics hardware and have high performance.}
\end{table*}

%% file: 3_dressi/dressi.tex
\section{Dressi}
\label{sec:dressi}

We depict the pipeline of our proposed Dressi in \figurename~\ref{fig:pipeline}.
Dressi-AD describes all the components of our DR algorithms.
Dressi takes inverse problems as inputs and outputs optimized results.
It consists of several components, such as Renderer, Optimizer, and Loss function.
The core component is Renderer, which applies HardSoftRas after scene updates (e.g., animation).
In the following subsections, we describe the goals and designs for establishing this pipeline. Then, we introduce the implementation and validation of our pipeline.
The implementation details are provided in the supplemental material.
\input{3_dressi/3_0_design_goals}
\input{3_dressi/3_1_system_design}
\input{3_dressi/3_2_0_AD_pipeline}

\input{3_dressi/3_2_1_function_generation}
\input{3_dressi/3_2_2_inverse_uv}
\input{3_dressi/3_2_3_stage_packing}
\input{3_dressi/3_2_4_gpu_execution}
\input{3_dressi/3_3_hardsoftras}

%% file: 3_dressi/3_0_design_goals.tex
\subsection{System Goals}
\label{subsec:design_goals}
%
This paper describes Dressi, a high-performance differentiable renderer for practical usage that supports a wide range of hardware.
In short, the proposed Dressi concept is {{``fully written in graphics pipeline-based AD for DR''}}.
As shown in Table~\ref{tab:system_comparison}, the existing rasterization-based DR frameworks use general-purpose AD and supplement DR-specific functions using handwritten modules in the GPGPU API.
This typical design is not scalable to various graphics hardware and can cause performance degradations owing to a lack of optimization across the entire system.
To solve these problems, we set the design goals of Dressi as follows:
%
\begin{itemize}
    \item G1: {\textbf{Hardware-agnostic}}.
          To democratize DR and explore the possibility of its practical application in everyday life, the DR system must be hardware-independent.
          People use various hardware, including low-end mobile devices, but existing DR systems do not support most of them.

    \item G2: {\textbf{Hardware-accelerated}}.
          Acceleration by graphics hardware is practically essential for handling massive DR computations.

    \item G3: {\textbf{Adaptivity}}.
          Automatic and dynamic adaptation to running hardware to boost performance is desirable for a practical DR system.
          Naive implementation with a static pipeline can cause deterioration of efficiency for some hardware.

    \item G4: {\textbf{Ease of modification}}.
          A practical DR system should be easily customized with APIs that support all primitive operations for DR.
          Unlike traditional renderers, many optimization settings, not just shaders, must be set for each DR application.
          DR algorithms should be easily customized as well to incorporate the latest DR techniques.

    \item G5: {\textbf{Independent}}.
          Practical DR systems should support independent edge devices without communicating with remote servers.
          For real-time applications with user interaction, the latency caused by communications should be avoided.
          Moreover, DR plays a vital role in appearance modeling using personal information (e.g., human face).
          Such modeling algorithms should be performed on edge devices with local data to protect user privacy.

    \item G6: {\textbf{Optimization friendly}}.
          Optimization with images is a primitive purpose of DR.
          DR should be capable of robust and fast optimization.
\end{itemize}

%% file: 3_dressi/3_1_system_design.tex
\subsection{System Design}
\label{subsec:system_design}
%
To achieve our goals, we make the following design choices:
\begin{itemize}
      \item C1: {\textbf{Graphics pipeline only}} ($\Rightarrow$ G1 and G2).
            We build all components of our system on a standard graphics pipeline.
            Consequently, our Dressi obtains cross-platform properties and acceleration by graphics hardware.

      \item C2: {\textbf{Monolithic system}} ($\Rightarrow$ G4 and G5).
            We design our AD and Dressi as a monolithic system.
            Considering only our AD, users can quickly develop DR applications and update Dressi itself.
            An independent device completes the DR execution.

      \item C3: {\textbf{Runtime optimization}} ($\Rightarrow$ G3).
            Hardware capability varies across devices (e.g., the number of input attachments of Vulkan).
            Our system dynamically generates and optimizes shader codes at runtime and maximizes the performance of each device.

      \item C4: {\textbf{Fully written in AD}} ($\Rightarrow$ G2, G3 and G4).
            Our DR algorithms are fully written in our AD for DR.
            Thus, the DR algorithms are wholly separated from the AD.
            The APIs of our AD provide all primitive operations for DR, such as rasterization and texture sampling.
            Our AD globally optimizes Dressi's performance.
            Users can extend Dressi simply to write forward passes with the functions of the AD.

      \item C5: {\textbf{Far-range gradient}} ($\Rightarrow$ G6).
            Our rendering process is designed to deliver far-range gradients from the screen space to vertex attributes.
            It enhances fast and robust convergence.
\end{itemize}

%
Based on the design choices above, Dressi is materialized with the following implementations:
\begin{itemize}
      \item I1: {\textbf{Dressi-AD} in \subsecname ~\ref{subsec:ad_pipeline}} ($\Rightarrow$ C1, C2, C3 and C4).
            A new AD library for DR is proposed.
            The backend of Dressi-AD is a cross-platform graphics pipeline API, Vulkan.
            Dressi-AD describes all components of Dressi; therefore, Dressi obtains hardware independence and acceleration.
      \item I2: {\textbf{Inverse UV} in \subsecname ~\ref{subsubsec:inverse_uv}} ($\Rightarrow$ C1 and C4).
            Regarding the backward computation of texture sampling with hardware in our AD, we propose an inverse UV texture.
      \item I3: {\textbf{Stage packing} in \subsecname ~\ref{subsubsec:stage_packing}} ($\Rightarrow$ C2 and C3).
            We propose a JIT compilation technique for Vulkan's hierarchy to map multiple subpasses to a render pass.
            Shader codes on the computational graph are automatically packed into optimal GPU execution units, considering hardware constraints.
            Reactive cache is integrated to skip redundant computation.
      \item I4: {\textbf{HardSoftRas} in \subsecname ~\ref{subsec:hardsoftras}} ($\Rightarrow$ C1 and C5).
            Contrary to most DR systems, ours is fully implemented under the limitation of a graphics pipeline.
            HardSoftRas is a novel rendering process enabling far-range gradient with graphics hardware.
\end{itemize}

%% file: 3_dressi/3_2_0_AD_pipeline.tex
\subsection{Dressi-AD: A Vulkan-Based AD Library for DR}
\label{subsec:ad_pipeline}
%
Dressi-AD, a hardware-agnostic AD for DR built on Vulkan, is the foundation of our system.
OpenGL was another candidate for the graphics pipeline, but it is too abstract to exploit the performance of modern hardware.
\diffred{OpenCL and compute shaders may not be supported well on mobile devices, and they cannot use hardware rasterizers.
    Moreover, their performances are limited since there is no correspondence to Vulkan's subpasses.}
We prefer Vulkan because of its low-level APIs and the potential to increase the rendering speed.
Unlike multi-dimensional tensors used in most existing AD libraries, we choose 2D images as the primitives of our AD, considering compatibility with graphics pipeline and DR.
Dressi-AD has all primitive operations for DR, including rasterization and texture sampling.
Dressi-AD is written in C++17.

%
Dressi-AD adopts its own variable and function objects for reverse-mode AD.
The APIs provided in general languages are the preferred design for developers \cite{mark2003cg}.
Developers write mathematical problems with the APIs, then Dressi-AD builds a computational graph and executes it on GPUs using the define-and-run scheme.
An example of a user program is shown in the supplemental material.
The bottom part of \figurename~\ref{fig:pipeline} is the build and execution pipeline of Dressi-AD after users define the problems.
In this subsection, we explain three key operations in the pipeline: function generation, stage packing, and GPU execution.
We use Vulkan terms to introduce a specific implementation.\footnote{
    {\texttt{VkImage}} is a structure that handles 2D arrays in Vulkan.
        {\texttt{VkRenderPass}} is a structure that contains subpasses, I/O structures, and dependencies between subpasses.
    The details are provided in \url{https://www.khronos.org/registry/vulkan/specs/1.2-extensions/man/html/}}

%% file: 3_dressi/3_2_1_function_generation.tex
\subsubsection{Function Generation}
\label{subsubsec:function_generation}
%
Dressi-AD instantiates function objects and variable objects in programs written by users.
Each forward function object has the GLSL representation that is compatible with a fragment shader,
and it has a method to generate backward function objects.
This GLSL code generation step roughly follows TensorFlow.js ~\cite{smilkov2019tensorflow}.
Some function objects for DR cannot be defined using the built-in GLSL functions.
One such exception is the backward functions of texture sampling, which will be explained in \subsubsecname ~\ref{subsubsec:inverse_uv}.

%
The variable object is a data structure for the inputs and outputs of the functions.
For reasons described later, we need to split the shaders and output the intermediate variables from them.
Vertex shaders cannot export intermediate variables, and compute shaders are not well optimized for graphics purposes.
Therefore, we prefer fragment shaders with dummy vertex shaders as much as possible, which are commonly used in deferred rendering.
In our implementation, we convert the variables into {\texttt{VkImage}}.
They can contain any 2D array resources, such as textures, vertex buffers, and blend weight matrices.
Note that the variables can represent higher dimensional tensors as stacked 2D arrays.

%% file: 3_dressi/3_2_2_inverse_uv.tex
\subsubsection{Inverse UV: Backward for Hardware Texture Sampler}
\label{subsubsec:inverse_uv}
%
Existing rasterization-based DR systems implement texture sampling with software, for instance, by tensor operation in general-purpose AD libraries \cite{ravi2020pytorch3d} and handwritten CUDA modules \cite{Laine2020diffrast}.
In contrast, our texture sampling is a function of our AD based on a graphics pipeline.
Thus, we are interested in directly handling textures of graphics pipeline.
We use {\texttt{texture()}}, which is a function for texture sampling in GLSL with hardware acceleration, as a forward function object to map texture space to screen space.
Forward texture sampling with linear interpolation generates a pixel value from four neighboring pixels.
In the backward pass, the gradients from the four neighbors should be summed up.
However, a naive implementation is difficult because the current APIs of Vulkan do not fully support atomic float add operations.

%
Therefore, we propose an inverse UV texture, the lookup table for converting gradients from the screen space into texture space.
Algorithm~\ref{alg:inverse_uv} calculates it in a compute shader in Vulkan.
The inverse UV texture can only maintain the last updated values for each texel, and a constant sampling order may lead to an imbalance.
Some texels kept in the inverse UV are updated in every frame, whereas others are never updated.
Consequently, artifacts may appear in an optimized texture.
To solve this imbalance, we disrupt the sampling order by adding quasi-random numbers generated from a Sobol sequence \cite{SOBOL196786} to $p^{ss}_{i}$ for each frame.
The gradient can be propagated into four texels during iterative optimization.
An inverse UV map that samples 3D scenes efficiently for ray tracing-based DR \cite{nimier2021material} was recently proposed.
In contrast, ours is designed to handle the backward pass with a hardware texture sampler.

%
\begin{figure}[!t]
    \begin{algorithm}[H]
        \caption{Calculating an inverse UV texture}
        \label{alg:inverse_uv}
        \begin{algorithmic}[1]
            \Require the size $(W_{ss}, H_{ss})$ of screen space, a rasterized UV image ${\mathbf I}_{uv} \in {\mathbf R}^{W_{ss}\times H_{ss} \times 2}$ in screen space, the size $(W_{tex}, H_{tex})$ of texture space
            \Ensure an inverse UV texture, ${\mathbf I}_{inv\_uv} \in {\mathbf R}^{W_{tex} \times H_{tex} \times 2}$
            \For{all pixel indices, $i = 1, 2, ..., W_{ss} \times H_{ss}$}
            \State calculate a pixel position in screen, $p^{ss}_i$
            \State fetch and normalize UV value, $p^{tex}$ at $p^{ss}_i$ from ${\mathbf I}_{uv}$
            \State Store $p^{ss}_i$ as a pixel value at $p^{tex}$ in ${\mathbf I}_{inv\_{uv}}$
            \EndFor
        \end{algorithmic}
    \end{algorithm}
\end{figure}
%
%

%% file: 3_dressi/3_2_3_stage_packing.tex
%
\BeginFigure
\centering
\includegraphics[width=\linewidth]{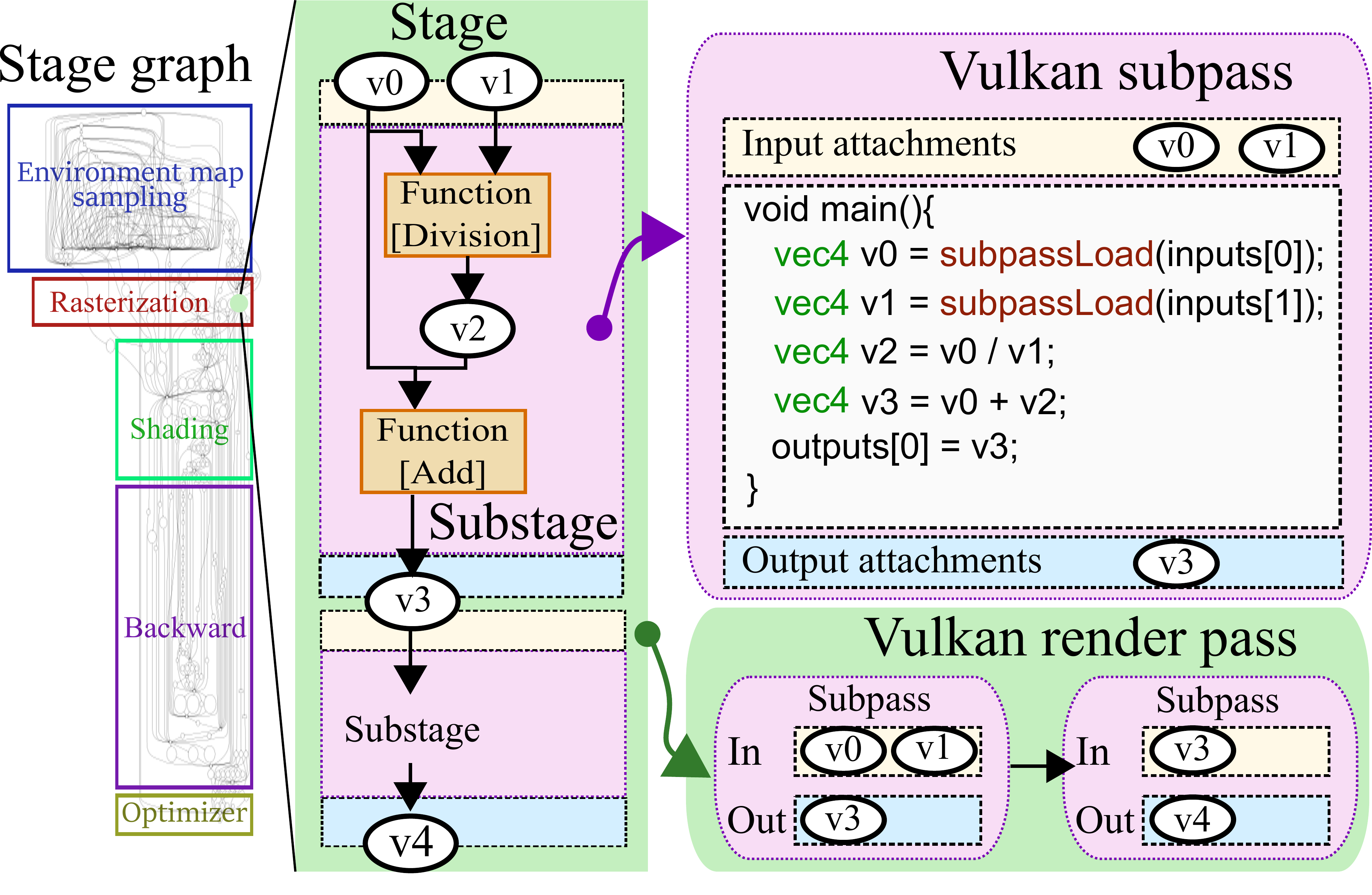}
\caption{Stage packing visualization.
  An example of a stage graph is shown on the left.
  A stage of rasterization is magnified from the middle to the right.
  Function objects are packed into a substage, and substages are packed into a stage.
    {\texttt{v0}}, ..., {\texttt{v4}} are variable objects.
  A substage corresponds to a subpass, and a stage is converted into a render pass in Vulkan.
  One subpass has one fragment shader, whose code is generated from the function objects in a substage.
  Variable objects, which are the input and output of a substage, are treated as attachments to a subpass.
  A render pass depends on input and output attachments among subpasses as Vulkan specification.
}
\label{fig:stage}
\EndFigure
%
%

%
\BeginFigure
\centering
\includegraphics[width=\linewidth]{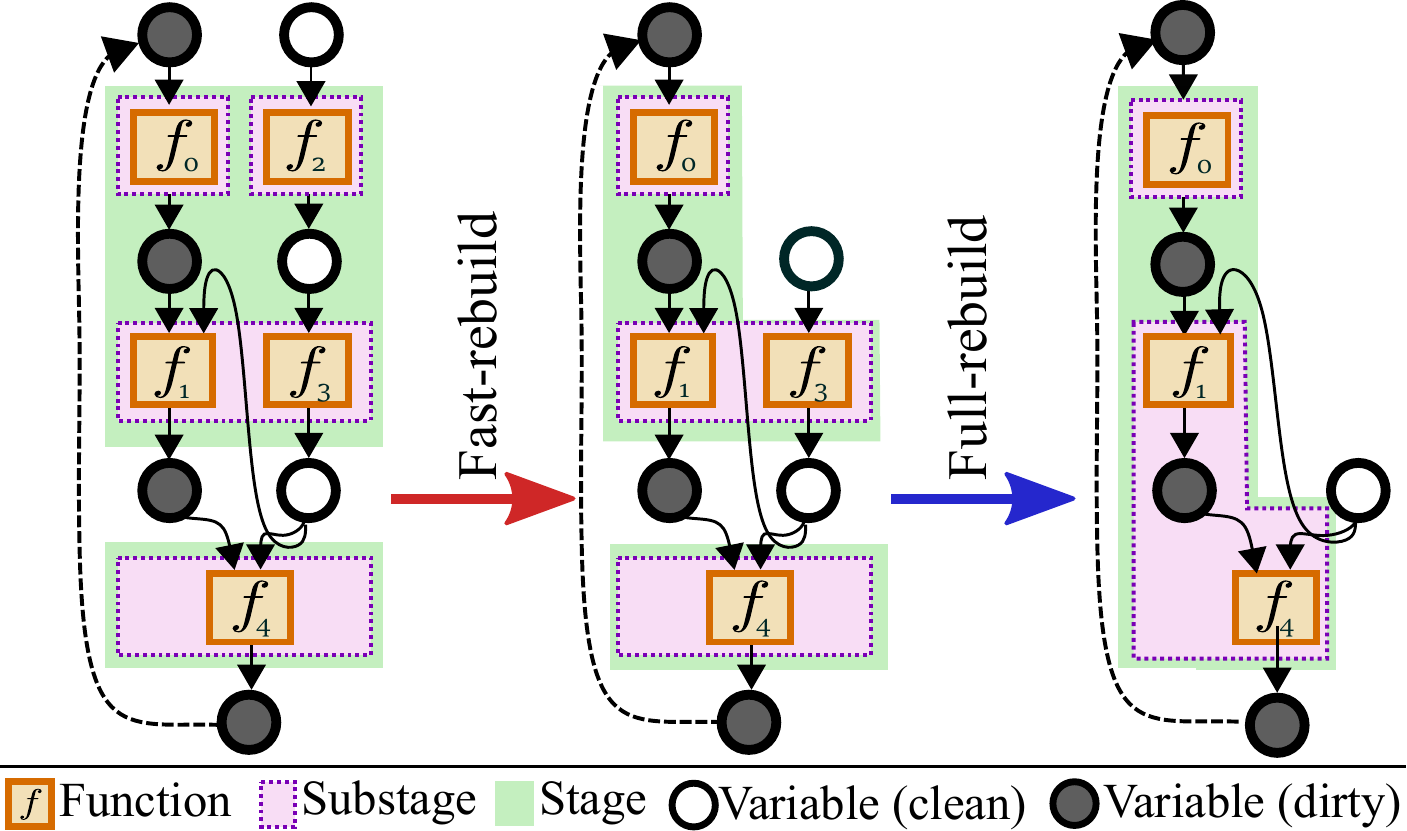}
\caption{Two-phased caching methods {\textit {fast-rebuild}} and {\textit {full-rebuild}} to use clean variable objects as cache.
  Fast-rebuild packs substages with dirty (i.e., not clean) variable objects as I/O into stages.
  Full-rebuild packs function objects with dirty variable objects as I/O into substages and constructs stages again.
  Full-rebuild is slower because it involves updating the shader code, but the graphs reconfigured with full-rebuild run faster than with fast-rebuild.
}
\label{fig:reactivity}
\EndFigure

\subsubsection{Stage Packing}
\label{subsubsec:stage_packing}
%
All variable objects and function objects for the forward pass, the backward pass, and optimizers to connect them constitute a single directed graph structure called a {\textit{computational graph}}.
The graph is not executable in a single fragment shader for the following reasons:
(1) Inverse UV computation by a compute shader and rasterization by a hardware rasterizer cannot be executed on fragment shaders.
(2) A single shader cannot have different sizes for the output images.
(3) There are limitations in the number of input and output variables in one shader.
These are typical limitations of graphics pipelines.
Therefore, we employ hierarchical shader packing to efficiently execute the graph in a reactive manner.

%
First, function objects in a computational graph are packed into {\textit{substage}}s, which represent fragment shaders, rasterizations, and compute shaders.
Each substage except a compute shader is compatible with a subpass of Vulkan.
Substages construct an oriented graph, {\textit{substage graph}}.
At the construction process, substage packing, a runtime optimization for each GPU model to minimize the number of substages with a greedy strategy, is performed.
Substage packing searches for a good combination of function objects to resolve hardware constraints and reduce bandwidth consumption.
As a result, it brings fewer read/write operations and improved memory efficiency.
Our method is inspired by the shader partitioning used in multipass rendering \cite{chan2002efficient}, which was proposed only for the forward pass.

%
After the construction of the substage graph, substages are merged into {\textit{stage}}s.
A {\textit{stage graph}} is built from the stages for execution as a render pass of Vulkan (i.e., {\texttt{VkRenderPass}}).
The two-level hierarchy between a render pass and multiple subpasses allows drivers to improve memory efficiency.
Although our graphs may contain compute shaders, Vulkan executes the compute shaders outside the hierarchy.
Vulkan has some constraints for render passes.
For example, subpasses in the same render pass must have the same image size.
Therefore, under these constraints, we apply stage packing to search for a better combination of substages.
Stage packing is done similarly as substage packing.
\figurename~\ref{fig:stage} shows the relationship among the function object, substage, and stage.

%
Our AD instantiates {\texttt{VkImage}} objects only from variable objects exposed beyond the substages, including all scene parameters ${\Theta}$.
Subpasses use the instances as input and output attachments of Vulkan.
Function objects in a substage are expanded into GLSL code by adding loading and saving {\texttt{VkImage}}.
As shader code simplification \cite{mcguire2006abstract}, we implement standard optimization techniques such as duplicated function removal and automatic reuse of {\texttt{VkImage}} to reduce memory consumption.
Additionally, a two-phased reactive cache is implemented to efficiently handle DR's complex graph structure and partially unchanged values on it.
Inspired by reactive programming \cite{Bainomugisha2013SurveyReactiveProgramming} and \diffred{sub-graph culling in a modern game engine \cite{yuriy2017framegraph},} our reactive cache automatically detects clean graph nodes that are not updated through optimization and skips their recalculation with cached values that have been computed once.
It is especially effective for large blocks such as irradiance sampling and rasterization for a static scene, although they can also work for small snippets.
\figurename~\ref{fig:reactivity} presents details of our reactive cache.
Although we have described the Vulkan implementation, our stage packing can be implemented for any graphics API with a hierarchical structure by minor modifications.
Our stage packing is a unique JIT compilation technique for DR because it (1) is designed for AD including backward, (2) handles the two-level hierarchy, and (3) has an embedded reactive cache.

%% file: 3_dressi/3_2_4_gpu_execution.tex
\subsubsection{GPU Execution}
\label{subsubsec:gpu_execution}
%
After stage packing, Vulkan objects are created according to a stage graph and substage graphs.
The order of stage execution is recorded into GPU command buffers.
Thanks to multiple subpasses in a render pass, Vulkan may automatically reduce bandwidth consumption, depending on the GPU vendor's implementation.

%% file: 3_dressi/3_3_hardsoftras.tex
\subsection{HardSoftRas: A Hardware Accelerated Soft Rasterizer}
\label{subsec:hardsoftras}
%
We propose a novel rendering process called HardSoftRas to propagate gradients from the screen space to far-range vertex attributes using hardware rasterizers.
Moreover, thanks to our full AD approach, the backward step is more efficient than prior art.
Dressi-AD can pack shader codes throughout HardSoftRas, although prior art cannot be optimized through common AD libraries and handwritten modules (see Table~\ref{tab:system_comparison}).
HardSoftRas consists of three main steps: rasterization, shading, and blending.

\subsubsection{Differentiable Rendering}
%
DR is a method that optimizes parameters of a 3D scene
by minimizing user-defined metrics between rendered images and ground truth (GT) images.
We formulate the scene parameters, ${ \Theta}$, with a set of 3D model, camera, ${\mathbf \theta}_{C}$,
lighting, ${\mathbf \theta}_{L}$,
and environment, ${\mathbf \theta}_{E}$, parameters.
The 3D model parameters are derived from the geometric, ${\mathbf \theta}_{G}$, and material, ${\mathbf \theta}_{M}$, parameters.
%
As Nvdiffrast indicates, rendering is regarded as a function composed of rasterization, shading, and post-processing, which can convert a 3D scene into 2D images.
We define the rendering function as follows:
\begin{align}
  {\mathbf  I}({ \Theta}) & = f_{render}({ \Theta})                                                        \\
                          & = f_{pp}(f_{shade}(f_{rasterize}({\mathbf  \theta}_{G},{\mathbf  \theta}_{C}),
  {\mathbf  \theta}_{M}, {\mathbf  \theta}_{L}, {\mathbf  \theta}_{E})).
\end{align}
where $f_{render}$, $f_{rasterize}$, $f_{shade}$, and $f_{pp}$ are rendering, rasterization, shading, and
post-processing (e.g., anti-aliasing, blending, gamma correction, and masking), respectively.
${\mathbf I}({ \Theta}) \in {\mathbf R}^{W \times H \times C}$ denotes the rendered image, where ${\mathbf  W}$, ${\mathbf  H}$, and ${\mathbf  C}$ are the width, height, and channel of the image.
To make the function differentiable with all parameters of the scene components,
the gradients of all the functions, $f_{rasterize}$, $f_{shade}$, and $f_{pp}$, should be calculated using the chain rule.
For example, given a loss function,
$L: {\mathbf  R}^{W \times H \times C} \times {\mathbf  R}^{W \times H \times C} \rightarrow {\mathbf  R}$,
to calculate the difference between the rendered, ${\mathbf I}({ \Theta})$, and reference, ${\mathbf  I}_{GT} \in {\mathbf  R}^{W \times H \times C}$ images.
An optimization problem for ${ \Theta}$ can be defined as
$\argmin_{{ \Theta}} L({\mathbf  I}({ \Theta}), {\mathbf  I}_{GT}).$
Then, the gradient of ${ \Theta}$
can be calculated as $\frac{\partial L}{\partial { \Theta}} = \frac{\partial L}{\partial {\mathbf  I}} \frac{\partial {\mathbf  I}}{\partial { \Theta}}.$
The optimization problem can be iteratively solved using a gradient-based method.

%
\begin{figure}[!t]
  \begin{algorithm}[H]
    \caption{Rasterization of HardSoftRas}
    \label{alg:hardsoftras_rasterize}
    \begin{algorithmic}[1]
      \Require The blur radius ${\mathbf r} \in [0, 1]$, \#buffers ${\mathbf K} \geq 1$, \#faces ${\mathbf F} \geq 1$, face indices $j \in {\mathbf F}$,  and faces ${\mathsf{f}}_j$
      \Ensure signed distance $dist^{k}_{i}$, depth value $depth^{k}_{i}$, and the other G Buffers $gb^{k}_{i}$ at each buffer, $k \in [1, {\mathbf K}]$ and pixel position $p_i$
      \For{ $k = 1, 2, ..., {\mathbf K}$}
      \For{all face indices $j = 1, 2, ..., {\mathbf F}$}
      \State ${\mathsf{f}}'_j \leftarrow {\mathsf{Enlarge}}({\mathsf{f}}_j, {\mathbf r})$
      \State Hardware rasterization to generate pixels on ${\mathsf{f}}'_j$
      \For{pixel indices $i \in {\mathsf{f}}'_j$}
      \State $dist_{ij} \leftarrow {\mathsf{SignedDist}}(p_i, {\mathsf{f}}_j)$
      \State Compute $depth'_{ij}$ and ${gb}_{ij}$
      \State $depth_{ij} \leftarrow {\mathsf{Shift}}(depth'_{ij}, dist_{ij})$
      \State Handle occlusion by depth peeling and depth test
      \State Update $dist^{k}_{i}$, $depth^{k}_{i}$, and $gb^{k}_{i}$
      \EndFor
      \EndFor
      \EndFor
    \end{algorithmic}
  \end{algorithm}
\end{figure}

\subsubsection{Rasterization}
%
The pixel-edge distance is essential for propagating far-range gradients.
SoftRas families handle correspondences between pixels and triangles in pixel-wise CUDA code.
Such complex pixel-wise processes cannot be implemented on graphics pipelines.
Based on the finding that the pixel-edge distance can be oppositely obtained from enlarged triangles, HardSoftRas utilizes face-wise calculations in the graphics pipeline.

%
The rasterization step of HardSoftRas is described in Algorithm~\ref{alg:hardsoftras_rasterize}.
The outer most loop is for depth peeling with ${\mathbf K}$ buffers.
The second outer loop with $j$ rasterizes faces ${\mathsf{f}}_j$ with geometry shaders, and the inner most loop updates the values at a pixel position, $p_i$, with index $i$ by fragment shaders.
Our key contributions are ${\mathsf{Enlarge}()}$, which enlarges the triangles in the screen space, and ${\mathsf{Shift}()}$, which updates the depth values of the enlarged face region ({\textit{soft face}}), considering the original face region ({\textit{hard face}}).

%
${\mathsf{Enlarge}()}$ makes the projected ${\mathsf{f}}_j$ larger to reach ${\mathbf r}$ from their edges in the geometry shaders.
${\mathbf r} \in [0, 1]$ defines the range of blurring in the screen space $\in [0, 1]^{2}$.
\figurename~\ref{fig:softras_rast} (a) depicts this process.
For acute projected triangles, we stretch their vertices in the opposite direction of their centroids (see \figurename~\ref{fig:softras_rast} (a) left).
However, this approach does not work well for obtuse triangles because the resulting triangle may become too sharp to cause shaggy boundaries.
Thus, we expand the bounding boxes to cover ${\mathbf r}$ if the triangles are obtuse.
The expanded bounding boxes are split into pairs of triangles for hardware rasterization (see \figurename~\ref{fig:softras_rast} (a) right).
This paper regards a triangle as obtuse if the smallest angle is less than 30\textdegree.
\diffred{Our strategy is less computationally expensive and more robust than the conservative rasterization \cite{hasselgren2005conservative} by keeping the number of triangle primitives as small as possible and avoiding precision errors of floating-point computation in the obtuse triangles.}
\figurename~\ref{fig:pseudo_occ} (a), (b), and (c) show how our ${\mathsf{Enlarge}()}$ works.
For $p_i$ on the expanded face, ${\mathsf{f}}'_j$, we compute the signed distances, $dist_{ij}$, from the original faces, ${\mathsf{f}}_j$, by ${\mathsf{SignedDist}()}$, as shown in \figurename~\ref{fig:softras_rast} (b).
These distances are used to deliver gradients in the screen space during the blending step.

%
If we expand a soft face by interpolated depth values from its hard face,
the soft face may cause artifacts that occlude the necessary hard faces around it.
We refer to this problem as {\textit{pseudo occlusion}} in this paper (see \figurename~\ref{fig:pseudo_occ} (d), for instance).
In fact, SoftRas and PyTorch3D also have pseudo occlusions; however, this is not a significant problem for them because their blending with numerous buffers (hundreds or thousands) cancels out the artifacts.
We want to keep ${\mathbf K}$ small at a maximum of five, due to the linear order of depth peeling.
To avoid pseudo occlusions with few buffers, we should assign more importance to hard faces than soft faces.
If $dist^{k}_{i} \geq 0$, $p_i$ at the $k$th buffer comes from the hard faces that are inside the original faces.
Otherwise, it is located on soft faces.
Hence, our depth modification, ${\mathsf{Shift}()}$, is defined as follows:
\begin{equation}
  \label{eq:dressi_hardsoft_depth}
  {\mathsf{Shift}}(depth, dist)  =
  \begin{cases}
    0.5 * depth        & {\mathrm{if}} \: dist \geq 0, \\
    0.5 * |dist| + 0.5 & {\mathrm{if}} \: dist < 0.
  \end{cases}
\end{equation}
With $depth \in [0, 1]$ and $dist \geq -1$, ${\mathsf{Shift}()}$ places all hard faces in front of all the soft faces.
Moreover, soft faces close to the hard faces with a high probability should be rasterized in the buffers.
To prioritize such soft faces, we update the depth values of soft faces based on $dist$.
\figurename~\ref{fig:softras_rast} (c) and the top row of \figurename~\ref{fig:pseudo_occ} show how ${\mathsf{Shift}()}$ works.
We describe the forward pass above.
We do not need special care for the backward pass, thanks to our full AD approach.

%
\BeginFigure
\begin{minipage}[b]{0.4\linewidth}
  \includegraphics[width=3.3cm]
  {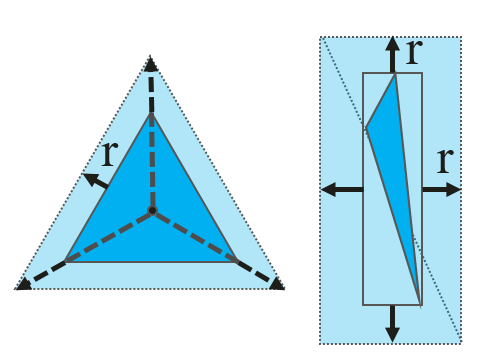}
  \subcaption{ ${\mathsf{Enlarge}()}$ }\label{fig:softras_enlarge}
\end{minipage}
\begin{minipage}[b]{0.25\linewidth}
  \centering
  \includegraphics[width=1.3cm]
  {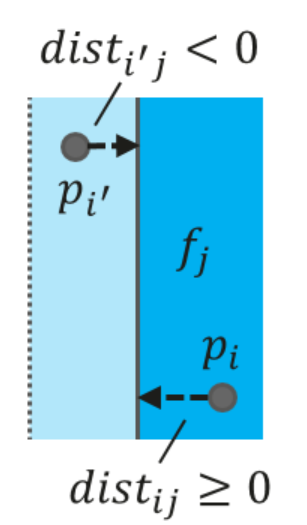}
  \subcaption{ ${\mathsf{SignedDist}()}$ }
  \label{fig:softras_dist}
\end{minipage}
\begin{minipage}[b]{0.25\linewidth}
  \centering
  \includegraphics[width=1.5cm]
  {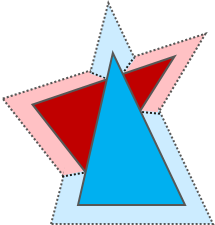}
  \subcaption{ ${\mathsf{Shift}()}$ }\label{fig:softras_modify}
\end{minipage}
\caption{
  Subroutines used in the rasterization step of HardSoftRas.
  Triangles in dark blue and red are original hard faces, and triangles in light blue and red are soft faces expanded from the hard faces, respectively.
  (a) Left: ${\mathsf{Enlarge}()}$ for acute triangles.
  Right: ${\mathsf{Enlarge}()}$ for obtuse triangle.
  (b) ${\mathsf{SignedDist}()}$ to compute the signed distance between the pixel position and the edge of a face.
  (c) Occlusion handling by ${\mathsf{Shift}()}$.
  The dark blue hard face is in front of the dark red hard face.
  Soft faces are pushed behind the hard faces, and those closer to the hard faces have closer depth values.}
\label{fig:softras_rast}
\EndFigure
%
%

%
\BeginFigure
\begin{minipage}[b]{0.24\linewidth}
  \center
  \includegraphics[width=1.7cm]{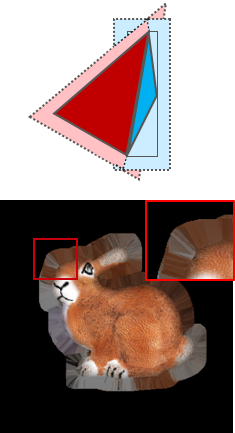}
  \subcaption{}
\end{minipage}
\begin{minipage}[b]{0.24\linewidth}
  \center
  \includegraphics[width=1.7cm]{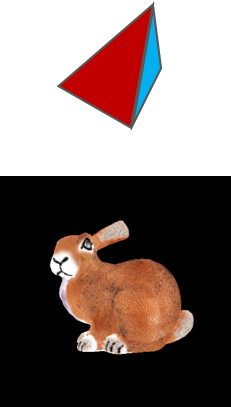}
  \subcaption{}
\end{minipage}
\begin{minipage}[b]{0.24\linewidth}
  \center
  \includegraphics[width=1.7cm]{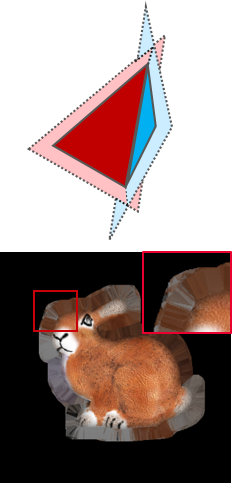}
  \subcaption{}
\end{minipage}
\begin{minipage}[b]{0.24\linewidth}
  \center
  \includegraphics[width=1.7cm]{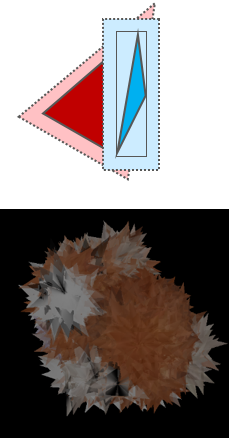}
  \subcaption{}
\end{minipage}
\caption{${\mathsf{Enlarge}()}$ and ${\mathsf{Shift}()}$ effects.
The top row shows the occlusion handling of two adjacent triangles, the blue triangle has a closer depth than red one, and the bottom row shows the blended images with ${\mathbf K}=2$.
(a) w/ ${\mathsf{Enlarge}()}$ and w/ ${\mathsf{Shift}()}$. Smoother soft faces than acute only, (c). This setting is the default.
(b) w/o ${\mathsf{Enlarge}()}$. Normally shaded image.
(c) w/ acute only ${\mathsf{Enlarge}()}$ and w/ ${\mathsf{Shift}()}$.
In the magnified red rectangle, boundaries of soft faces are shaggy.
(d) w/ ${\mathsf{Enlarge}()}$ but w/o ${\mathsf{Shift}()}$. Suffering from pseudo-occlusions. In this case, we visualize all faces without an edge mask. }
\label{fig:pseudo_occ}
\EndFigure

\subsubsection{Shading and Blending}
%
After rasterization, deferred shading is applied to each G buffer $gb^{k}_{i}$ to generate a shaded buffer.
Dressi supports a variety of shadings such as physically-based shading (PBS), image-based lighting (IBL), skin, and shadows.

%
Finally, we blend the shaded buffers per pixel to determine the final pixel colors of a shaded image.
Our blending is based on SoftRas but we update it in several aspects.
First, we calculate the probability
$\mathcal{D}_k = {\mathsf{sigmoid}}(dist_k/\sigma), \sigma>0$, at $p_i$ of the $k$th buffer.
We use $\sigma = {\mathbf r} / 7$, unless otherwise noted.
To deal with the numerical instability of sigmoid computation, we remove the square for $dist_k^i$ in SoftRas.

%
Our blending function is generally composed similarly to traditional alpha blending as follows:
\begin{equation}
  I(I^{e}, I^{ne}) = E I^{e} + (1-E) I^{ne}.
\end{equation}
where $I$ denotes a pixel value at $p_i$, and $E$ is a binary edge mask.
$I^{e}$ and $I^{ne}$ are the blended pixel values for the edges and non-edges, respectively.
Our formulation explicitly distinguishes blending for edges,
which involve many vertical surfaces from a view and have a strong signal for optimization. Thus, $I^{e}$ should have transparency and far-range gradients.
In contrast, non-edges have a weak signal, and transparency for $I^{ne}$ is not necessary in most cases.
We compute $E$ from $S^H$, which is a stencil mask of hard faces in the frontal buffer by edge detection and dilation.
Therefore, its width, $\delta \leq {\mathbf r}$, of $E$ is adjustable.
We set $\delta = {\mathbf r}$, unless otherwise noted.

%
To compute a color pixel value, $I_c$,
we use weighted averaging for edges, $I_{c}^{e}$, and keep the most frontal shaded color, $C_1$, for non-edges, $I_{c}^{ne}$.
\begin{equation}
  I_c= I(I_{c}^{e}, I_{c}^{ne}),\; I_{c}^{e}= \frac{ \sum_{k}{\mathcal{D}_k C_k} }{ \sum_{k}{\mathcal{D}_k} },\; I_{c}^{ne} = S^H C_1.
\end{equation}
where $C_k$ denotes the shaded color of the $k$th buffer.
We do not use depth values for our color blending as SoftRas because transparency tuning is difficult.

%
For a silhouette pixel value, $I_s$, we use binary occupancy blending for edges, $I_{s}^{e}$, and flat intensity for non-edges, $I_s^{ne}$.
\begin{equation}
  I_s=I(I_{s}^{e}, I_{s}^{ne}),\; I_s^e = 1 - \prod_{k}(1 - \mathcal{D}_k),\; I_s^{ne} = S^H.
\end{equation}

%
Our formulation above generalizes and bridges anti-aliasing and SoftRas.
A comparison of HardSoftRas with some settings and the existing DR techniques is shown in \figurename~\ref{fig:softras_generalize}.
For example, Nvdiffrast is inspired by distance-to-edge AA \cite{malan2018edge} and geometric post-process AA \cite{GPAA2011}.
It detects edges in the screen space geometrically, computes pixel-to-edge distances for pixels closer to the edges of less than one pixel in the most frontal buffer, and blends colors for them.
When we set $\delta$ to one pixel equivalent with ${\mathbf K}=1$,
our method can approximate Nvdiffrast with a subtle visual difference ((a) and (d)).
One difference is that we blur all edge pixels, whereas Nvdiffrast selects a part of the edge pixels to be blended.
Moreover, our edge detection employs image processing for silhouette edges, but Nvdiffrast's geometric method may detect inside edges on depth discontinuity.
On the other hand, SoftRas and PyTorch3D blend the non-edge region besides the edge region.
By setting $E(p_i)=1 | {p_i\in \forall i}$, our formulation approaches them by blending faces inside silhouettes ((c), (e), and (f)).
Our appearance is slightly different, owing to the updated blending function and edge-pixel distance computing region of the face-wise approach.
Our moderate and preferred setting (b) with ${\mathbf r}=0.01, {\mathbf K}=3$ maintains a sharp texture inside silhouettes and generates far-range gradients around the edges.
\diffred{In general, far-range settings like SoftRas converge faster, while near-range settings like AA are useful to accurately optimize complex scenes with many overlaps.}
\diffred{A recent optimization technique \cite{Nicolet2021Large} can be combined with our method.}

%
\BeginFigure
\centering
\begin{minipage}[t]{0.32\linewidth}
  \centering
  \includegraphics[width=2.0cm]{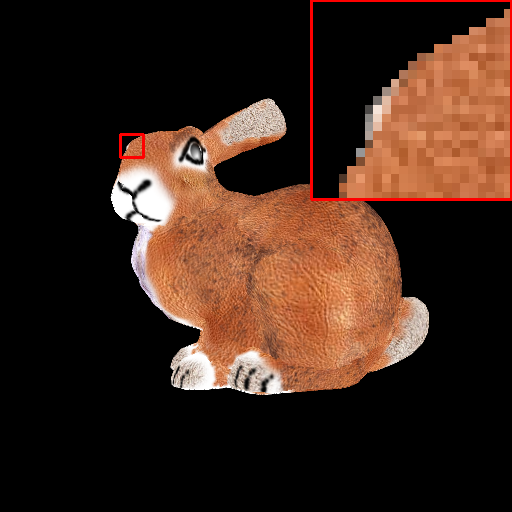}
  \subcaption{\textbf{Ours \scriptsize{(AA mode)}}}
\end{minipage}
\begin{minipage}[t]{0.32\linewidth}
  \centering
  \includegraphics[width=2.0cm]{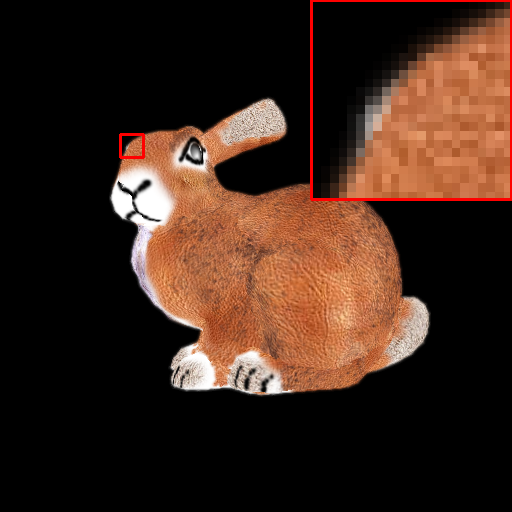}
  \subcaption{\textbf{Ours}}
\end{minipage}
\begin{minipage}[t]{0.32\linewidth}
  \centering
  \includegraphics[width=2.0cm]{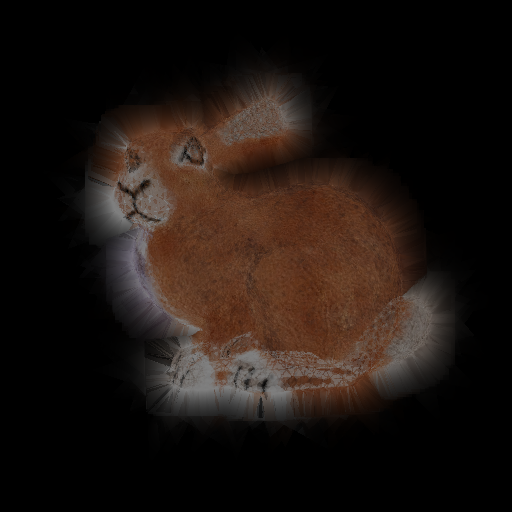}
  \subcaption{\textbf{Ours \scriptsize{(SoftRas mode)}}}
\end{minipage}
\\
\begin{minipage}[t]{0.32\linewidth}
  \centering
  \includegraphics[width=2.0cm]{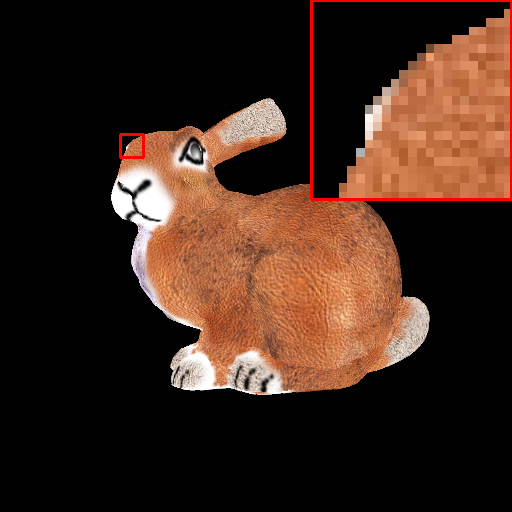}
  \subcaption{Nvdiffrast}
\end{minipage}
\begin{minipage}[t]{0.32\linewidth}
  \centering
  \includegraphics[width=2.0cm]{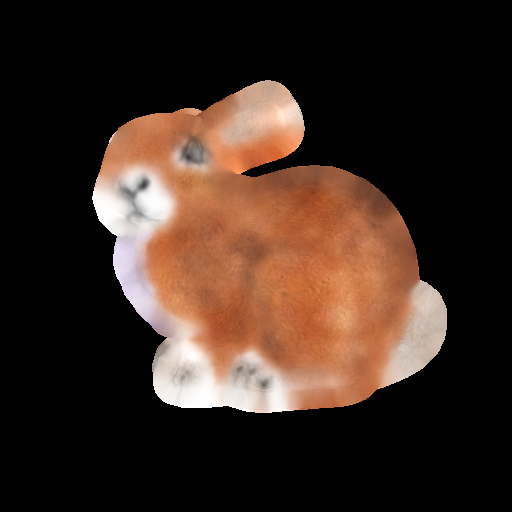}
  \subcaption{PyTorch3D \scriptsize{(\#buf=150)}}
\end{minipage}
\begin{minipage}[t]{0.32\linewidth}
  \centering
  \includegraphics[width=2.0cm]{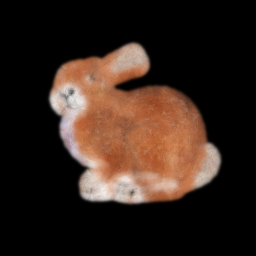}
  \subcaption{SoftRas}
\end{minipage}
\caption{Comparison of HardSoftRas with several settings and existing methods with bunny model (\#face=5k) of image size 512 $\times$ 512.
  We use $\mathbf{r}=0.1, \sigma=0.0285$ for (c) and set the corresponding parameters to (e) and (f).
  Red rectangles for (a), (b), and (d) show blurring range differences.
}
\label{fig:softras_generalize}
\EndFigure
%
%

%% file: 4_validation/validation.tex
%
\BeginFigure
\center
\includegraphics[width=6.0cm]{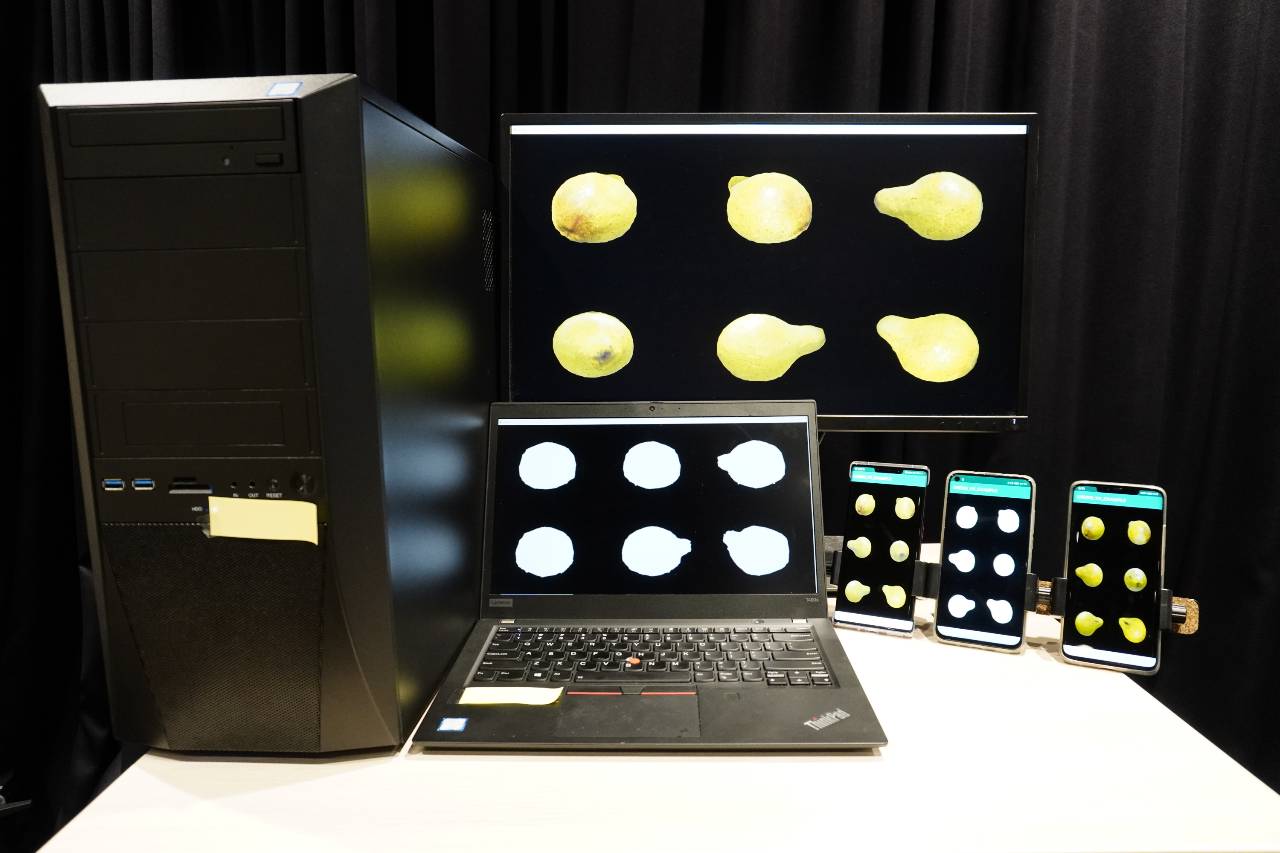}
\caption{
  Cross-platform DR application running on various devices:
  a desktop PC with NVIDIA, laptop with Intel, and smartphones with Arm.}
\label{fig:demo_3_platforms}
\EndFigure
%
%

%
\BeginFigure
\begin{minipage}[b]{0.75\linewidth}
  \includegraphics[width=5.63cm]{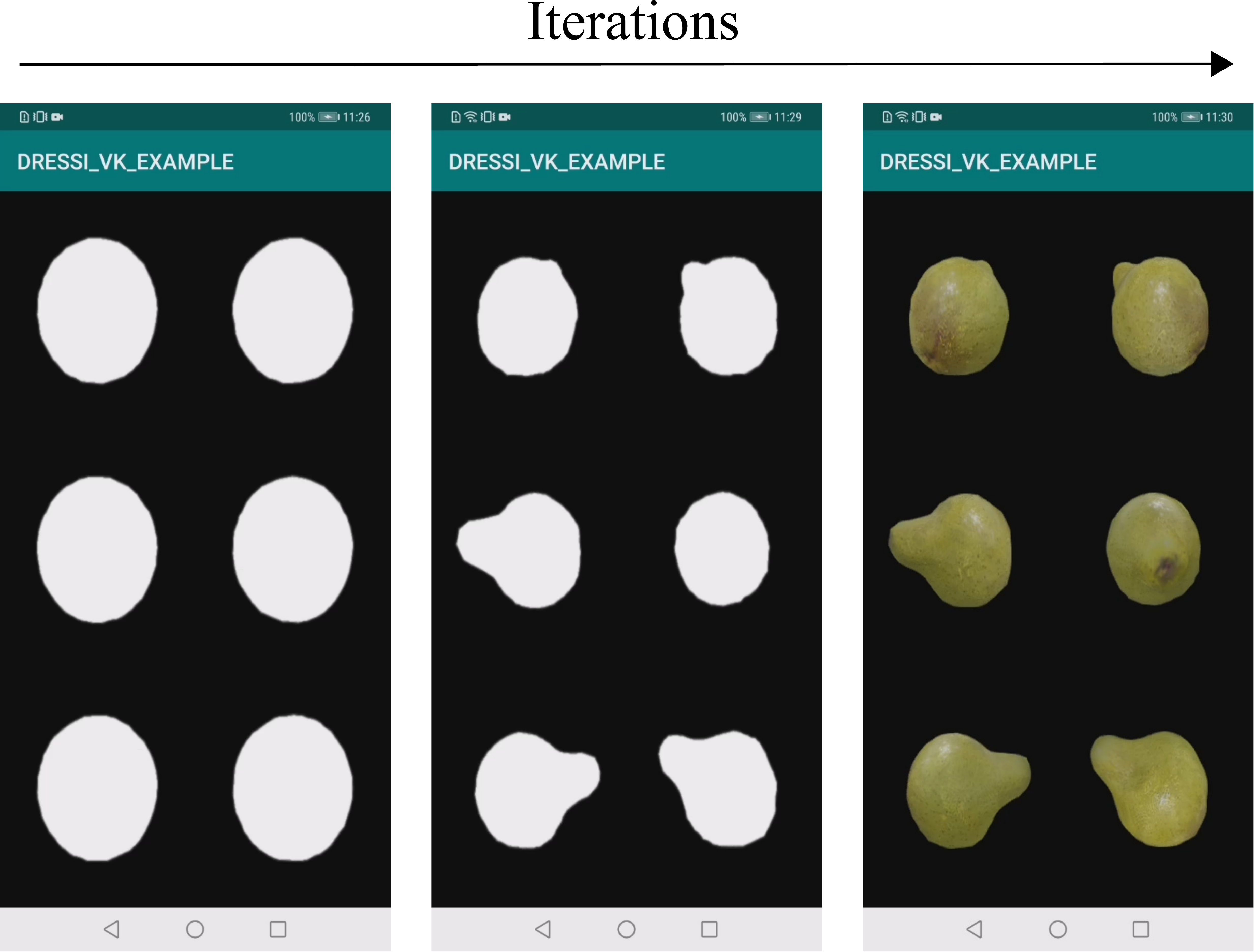}
  \subcaption{Optimization process}
\end{minipage}
\begin{minipage}[b]{0.22\linewidth}
  \includegraphics[width=1.75cm]{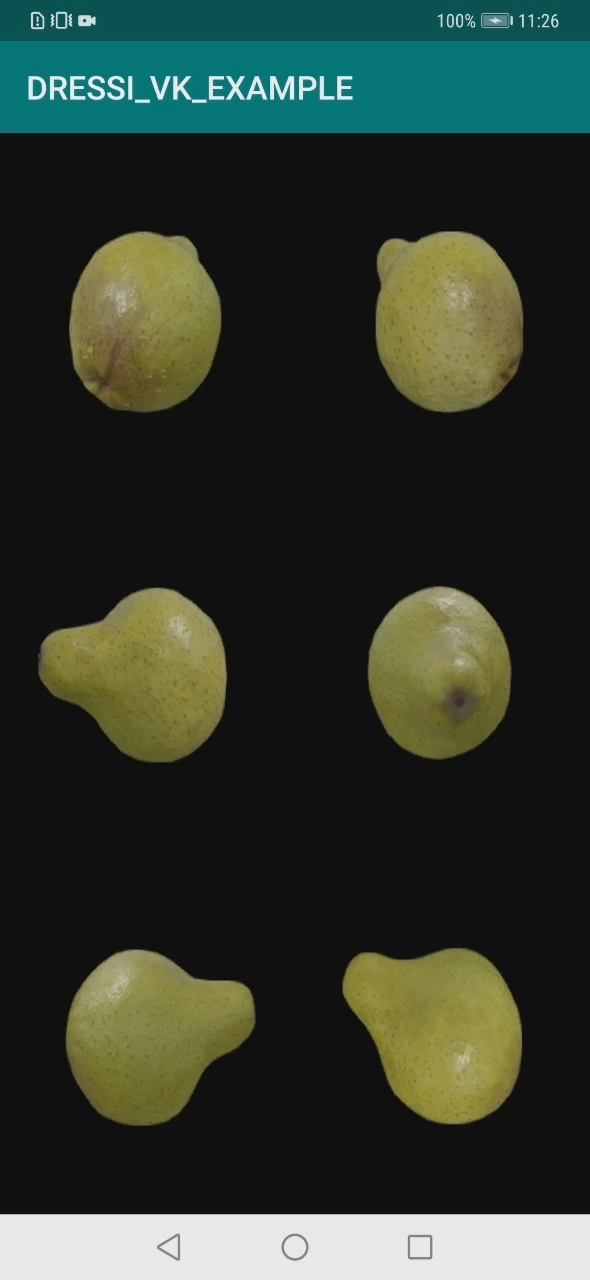}
  \subcaption{GT}
\end{minipage}
\caption{
  Cross-platform DR application on a mobile platform.
  It optimizes the vertex positions and albedo texture of a white sphere to fit the rendered result to the green pear images.
  (a) Optimization process.
  From left to right: rendered images with an initial sphere geometry with a white albedo texture, optimized geometry, and optimized geometry and texture.
  (b) Rendered GT pear model.}
\label{fig:mobile_app}
\EndFigure
%
%

%
\begin{table*}[t]
  \center
  \begin{tabular}{ccccc|cc|cc}
                                            &                                          &                                          &                        &                           & \multicolumn{2}{c}{\footnotesize{\#Substage / \#Stage (Before $\rightarrow$ After cache)}} & \multicolumn{2}{l}{\footnotesize{Time per iter. (ms)}}                                                       \\\hline
    \multicolumn{1}{c}{Vendor}              & \multicolumn{1}{c}{\footnotesize{Model}} & \multicolumn{1}{c}{\footnotesize{Usage}} & \footnotesize{Used OS} & \begin{tabular}{c}
      \footnotesize{Maximum} \\ \footnotesize{\#input attach.}
    \end{tabular} & \footnotesize{Geo. opt.}                                                                   & \footnotesize{Tex. opt.}                               & \footnotesize{Geo. opt.} & \footnotesize{Tex. opt.} \\ \hline \hline
                                            & \scriptsize{RTX2080}                     & \scriptsize{Workstation}                 & \scriptsize{Windows}   & \footnotesize{1048576}    & \footnotesize{218/129 $\rightarrow$ 206/122}                                               & \footnotesize{187/160 $\rightarrow$ 86/69}             & \footnotesize{3.0}       & \footnotesize{2.0}       \\
    \multirow{-2}{*}{\footnotesize{NVIDIA}} & \scriptsize{TelsaT4}                     & \scriptsize{Server}                      & \scriptsize{Ubuntu}    & \footnotesize{1048576}    & \footnotesize{220/131 $\rightarrow$ 207/124}                                               & \footnotesize{186/160 $\rightarrow$ 88/70}             & \footnotesize{6.1}       & \footnotesize{4.1}       \\
    \footnotesize{AMD }                     & \scriptsize{RadeonR9M360}                & \scriptsize{Desktop}                     & \scriptsize{Windows}   & \footnotesize{N.A.}       & \footnotesize{219/129 $\rightarrow$ 206/122}                                               & \footnotesize{187/160 $\rightarrow$ 86/69}             & \footnotesize{51.4}      & \footnotesize{28.8}      \\
    \footnotesize{Intel}                    & \scriptsize{UHDGraphics620}              & \scriptsize{Laptop}                      & \scriptsize{Windows}   & \footnotesize{8}          & \footnotesize{348/134 $\rightarrow$ 336/128}                                               & \footnotesize{212/181 $\rightarrow$ 111/84}            & \footnotesize{38.3}      & \footnotesize{23.5}      \\
    \footnotesize{Arm}                      & \scriptsize{Mali-G76}                    & \scriptsize{Smartphone}                  & \scriptsize{Android}   & \footnotesize{4}          & \footnotesize{687/142 $\rightarrow$ 670/132}                                               & \footnotesize{278/187 $\rightarrow$ 192/86}            & \footnotesize{420.0}     & \footnotesize{189.5}     \\ \hline
  \end{tabular}
  \caption{
    Stage packing and performance evaluation of the cross-platform DR application on various hardware.
    As an example of the hardware constraints for our stage packing, we show the maximum number of input attachments taken from the Vulkan API.
    The larger the value of the input attachment is, the fewer the number of stages, and the more efficient the execution.
    The hardware constraint is a concept orthogonal to the hardware performance (e.g., clock).
    The time taken for optimization depends on both hardware constraints and performance.
  }
  \label{tab:cross_platform_pf}
\end{table*}

\section{Validation}
\label{subsec:validation}

\subsection{Cross-Platform Validation}
\label{subsec:mobile_app}
%
To demonstrate the hardware independence, we ran the same geometry and texture optimization application on four major platforms: NVIDIA for workstations and servers, AMD for desktops, Intel for laptop PCs, and Arm for smartphones.
\figurename~\ref{fig:demo_3_platforms} depicts the application running on various hardware.
The application optimizes an initial sphere with a white albedo texture, bringing images rendered in six views closer to synthetic pear images.
Rendering image size per view and albedo texture size are $256 \times 256$.
First, the application optimizes the vertex positions by silhouettes with the L2 loss and Laplacian regularization.
After the geometry is aligned to the pear silhouettes, we optimize the diffuse albedo texture, minimizing color L2 loss.
We use the Adam optimizer for both optimization processes, and the camera parameters are fixed as the ground truth.
The same parameters are used for all the platforms.
\figurename~\ref{fig:mobile_app} shows screenshots of the application on a mobile platform with Arm.

%
Table~\ref{tab:cross_platform_pf} shows our detailed experimental settings, stage packing effects and performance.
The maximum number of input attachments for each hardware type, obtained from the Vulkan API, is a constraint used for stage packing.
Although more constraints are considered, we omit other constraints for need of space.
Our stage packing faithfully generates optimized stages for each platform: fewer stages for less constraint hardware (i.e., NVIDIA and AMD).
Our reactive cache reduces more stages for texture optimization by skipping constant rasterization results.
The two rightmost columns are the average optimization time per iteration of the platforms.
NVIDIA RTX2080 for gaming workstations shows the best number.
TeslaT4 for deep learning servers is slightly slower than the RTX2080 because it is not optimized for graphics purposes.
The second is an Intel for laptops, the third is an AMD for office desktops, and the last is an Arm for smartphones.
The stages reduced by our reactive cache make texture optimization faster than geometry optimization in all the platforms.

%
\figurename~\ref{fig:cross_platform_loss_eval} shows the error curves against GT.
We take the averages of over 10 trials to handle the non-deterministic behaviors of GPUs.
For geometry optimization, we show 3D distance error curves against GT geometry.
They converge at approximately 150 iterations on all platforms.
Converged error values are very similar for all platforms.
For texture optimization, all platforms converge around 50 iterations.
The AMD shows larger values than others in the middle of geometry optimization.
Although graphics pipeline API is common, its implementation (e.g., floating-point precision) depends on vendors.
Therefore, optimization results may have a little numerical difference among hardware.

%
\BeginFigure
\begin{minipage}[b]{0.5\linewidth}
  \includegraphics[width=3.9cm]{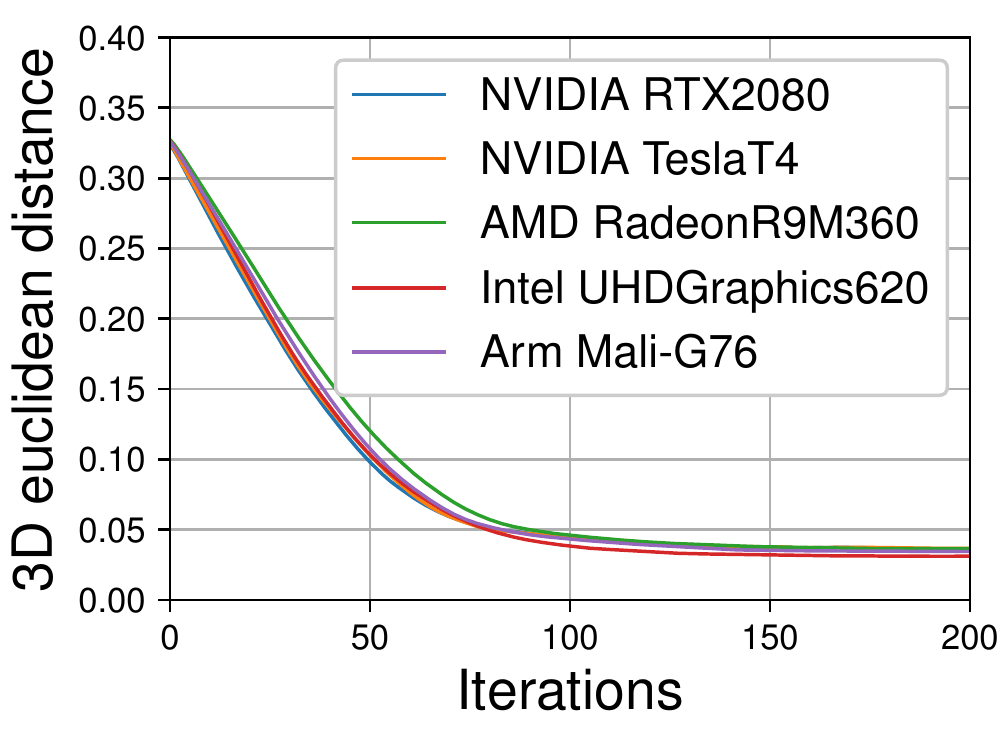}
  \subcaption{\small Geometry optimization}
\end{minipage}
\begin{minipage}[b]{0.5\linewidth}
  \includegraphics[width=3.9cm]{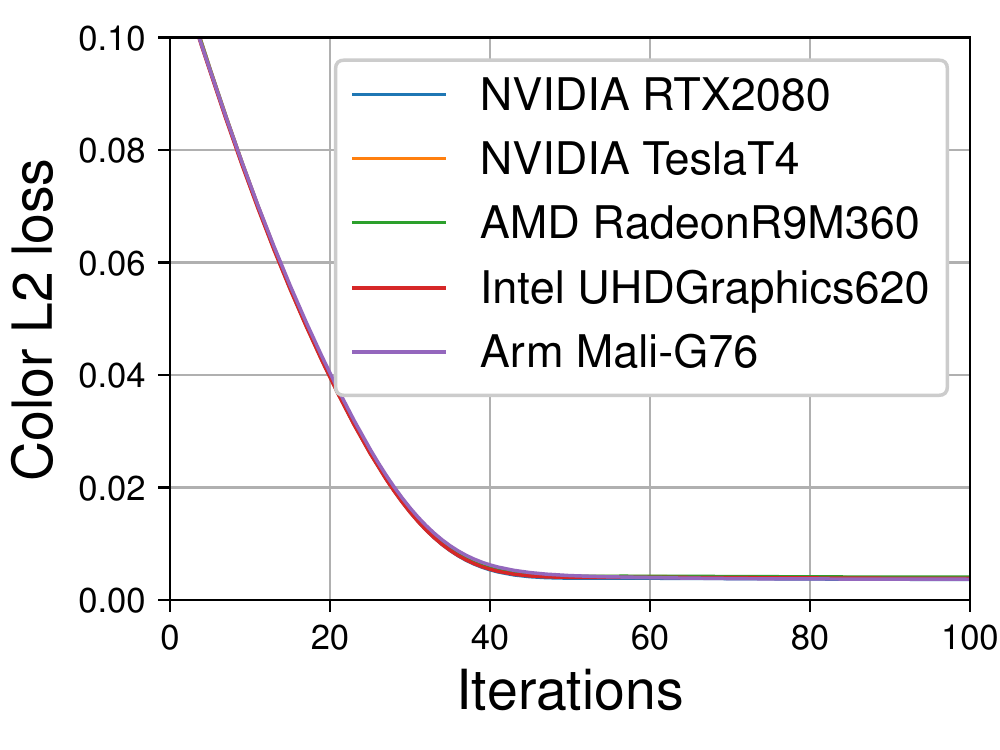}
  \subcaption{\small Texture optimization}
\end{minipage}
\caption{
  Error against GT on several platforms.
  Errors converge to almost the same value for all platforms, which indicates that our method is hardware-agnostic.
  (a) 3D distance (not used as a loss) between GT and optimized geometry in geometry optimization.
  (b) Color L2 loss during texture optimization.}
\label{fig:cross_platform_loss_eval}
\EndFigure
%
%

%
\input{supplements/stagepack_react_speed/table}

\subsection{Stage Packing Validation}
\label{subsec:stage_packing_validation}
%
We performed an ablation study to validate that our stage packing and reactive cache contribute to the fast speed.
As a baseline of stage packing off, we used a naive method to assign each function object to a single stage without the greedy strategy.

%
We performed the validation on Arm Mali-G76 for two reasons.
First, it is designed for mobile devices and has limited hardware capability.
Dressi should accelerate the DR on such weak hardware.
Second, it is based on a tile-based GPU architecture and suitable for confirming speed improvement by substages and the corresponding subpasses.
A subpass in Vulkan is designed for tile-based rendering, and its proper utilization reduces bandwidth use \cite{arm2021tile}.
Table~\ref{tab:stage_packing_validation} shows the comparison with several settings.
Our proposed approaches successfully accelerate forward and backward rendering speeds.

\subsection{HardSoftRas Validation}
\label{subsec:exp_synthetic}
%
We analyze our optimization capability with several sets of hyperparameters.
We optimize vertex positions of a coarse sphere geometry to make it visually similar to a dense synthetic GT model, which is a skull geometry with a complex and uneven shape.
At each iteration, a random view with a random point light in Lambertian shading is rendered.
The L1 loss for the rendered GT with the same configuration is minimized with the scheduled Laplacian regularization term.
We ran 2,500 iterations in 256 $\times$ 256 image resolution.
As a reference, we compared ours with Nvdiffrast, using the same parameters and settings.
We ran this experiment on a desktop PC with NVIDIA RTX2080, and the average 3D distance of vertices to the closest GT surface was used as an evaluation metric.

%
\figurename~\ref{fig:softras_cg_heatmap} shows the validation as a heatmap with $\mathbf{K}$ and $\mathbf{r}$ combinations.
At 500 iterations (a), wide blurring settings with large $\mathbf{K}$ and $\mathbf{r}$ show better results.
In contrast, at 2,500 iterations (b), a small $\mathbf{r}$ yields a better convergence.
The results show that far-range gradients are preferred at the early stages of optimization because the object being optimized is far from the target.
However, at the latter stages of the process, it is better to use near-range gradients to refine the details.

%
Next, we compare the error curves of our method and Nvdiffrast in \figurename~\ref{fig:softras_cg_loss}.
For our method, we plotted three configurations based on the validation heatmaps: (1) the best parameter at iteration 2,500 ($\mathbf{r}=2\mathrm{pix}, \mathbf{K}=3$), (2) parameters for a very far-range gradient ($\mathbf{r}=7\mathrm{pix}, \mathbf{K}=5$), and (3) parameters for a very near-range gradient ($\mathbf{r}=1\mathrm{pix}, \mathbf{K}=1$).
The error curves show the same tendency as the heatmaps.
Our convergence is faster in all the settings than Nvdiffrast.
Regarding accuracy, the final distance in 2,500 iterations, excluding the very far-range setting, is better than that of Nvdiffrast.

%
Finally, we apply normal map optimization to the optimized geometries to refine the appearance closer to the GT.
We used almost the same setting for geometry optimization but increased the size of rendered images to 1024 $\times$ 1024.
The size of a normal map was also set to 1024 $\times$ 1024.
We visually compare our method with the configuration (1) and Nvdiffrast in \figurename~\ref{fig:softras_cg_visual}.
Our method shows better convergence to the GT with a closed ring shape on the side.

%
\BeginFigure
\centering
\begin{minipage}[t]{0.45\linewidth}
  \centering
  \includegraphics[width=4.0cm]{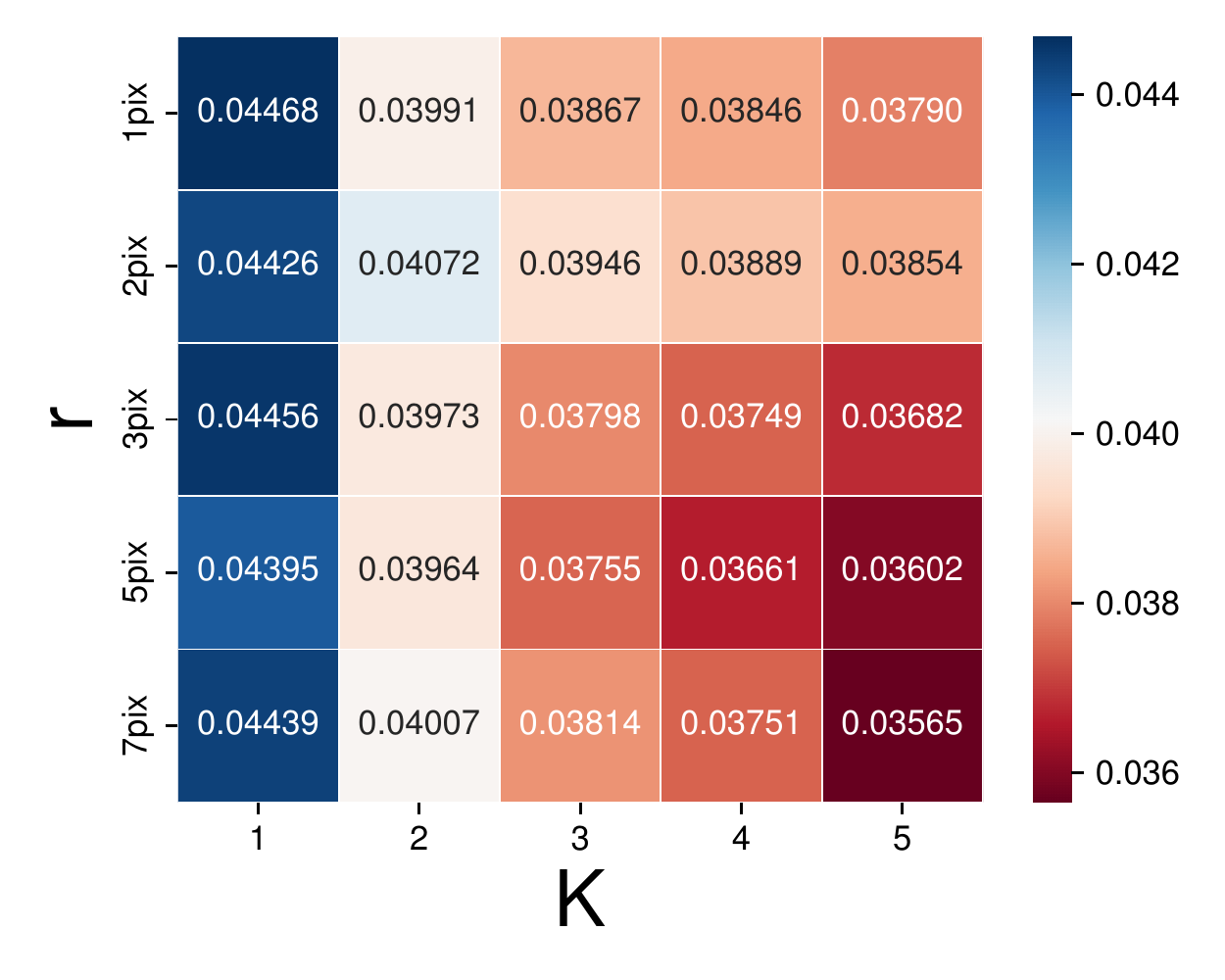}
  \subcaption{At iter. 500}
\end{minipage}
\begin{minipage}[t]{0.45\linewidth}
  \centering
  \includegraphics[width=4.0cm]{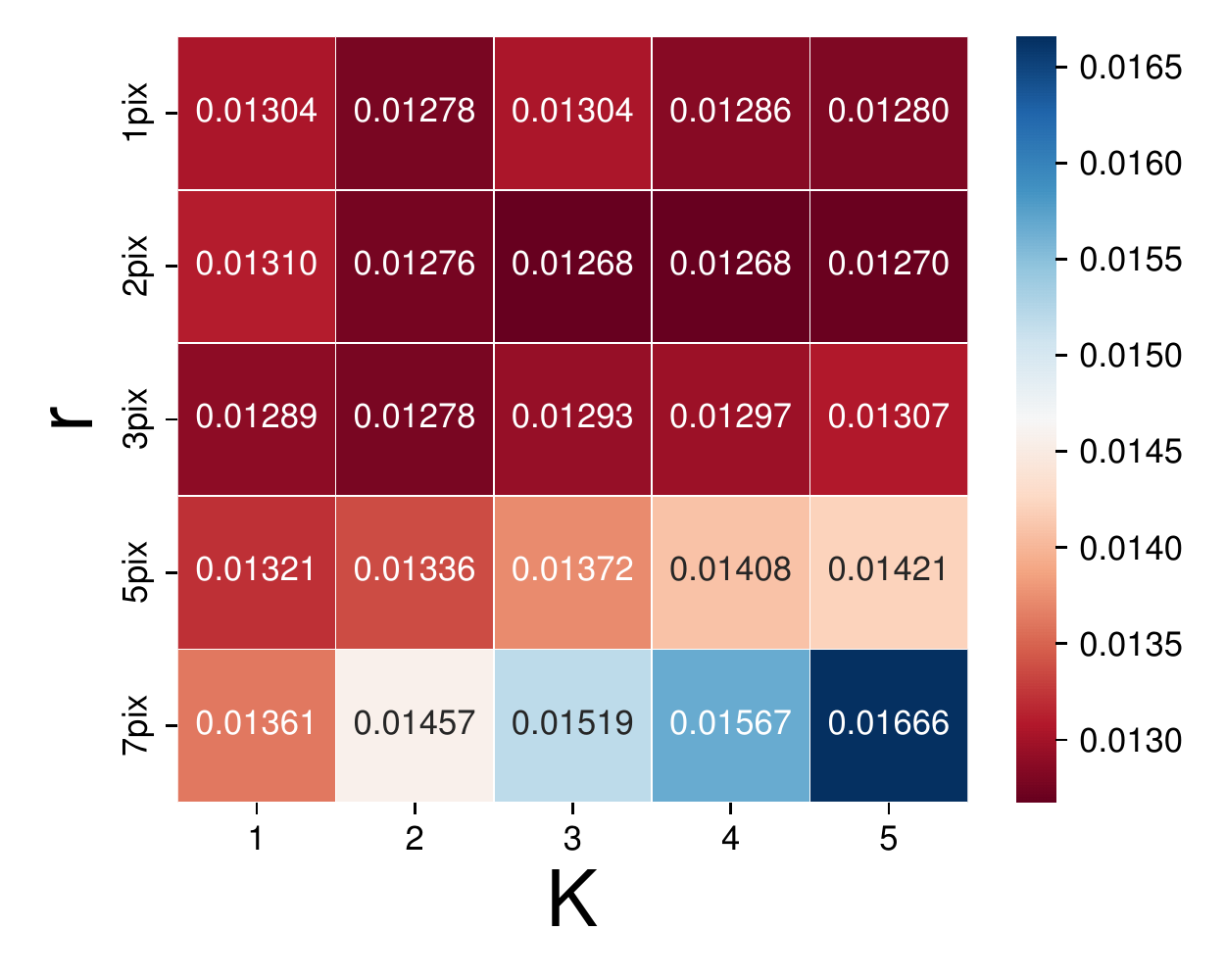}
  \subcaption{At iter. 2,500}
\end{minipage}
\caption{Validation heatmaps for hyper parameters.
  The 3D distances against the GT with several $\mathbf{K}$ (horizontal) and $\mathbf{r}$ (vertical) combinations are shown.
  The upper left with small $\mathbf{r}$ and $\mathbf{K}$ is a near-range gradient setting, and the lower right with large $\mathbf{r}$ and $\mathbf{K}$ is for far-range gradients.
  (a) After 500 iterations. Settings with larger $\mathbf{r}$ and larger $\mathbf{K}$ show better distances.
  (b) After 2,500 iterations. Settings with smaller $\mathbf{r}$ show better distances.}
\label{fig:softras_cg_heatmap}
\EndFigure
%
%

%
\BeginFigure
\centering
\includegraphics[width=8.0cm]{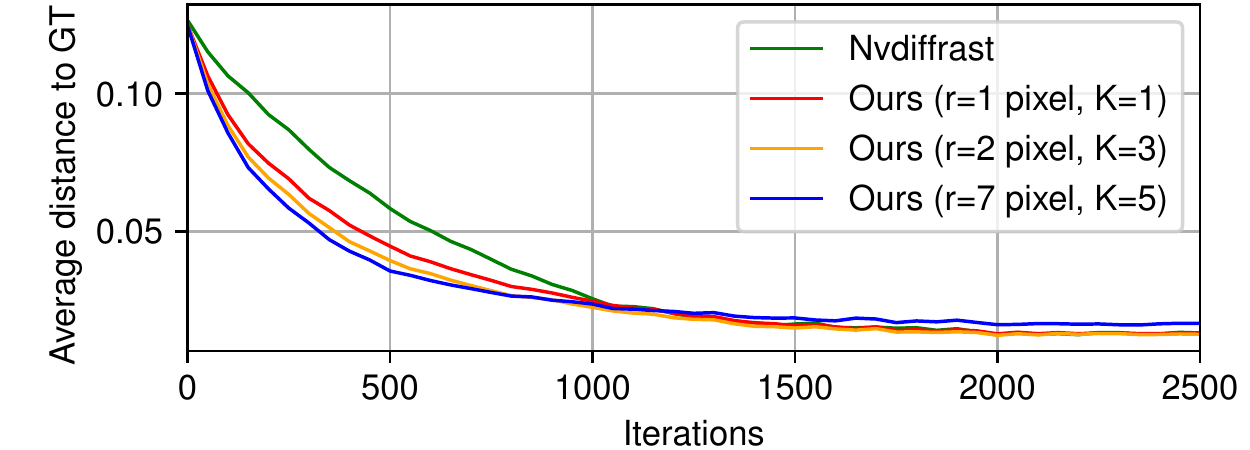}
\caption{
  3D distance against GT per iteration of ours with selected parameters and Nvdiffrast.
  Ours shows similar tendency to the heatmap and shows faster convergence than Nvdiffrast. }
\label{fig:softras_cg_loss}
\EndFigure
%
%

%
\begin{figure}[t]
\centering
\begin{minipage}[t]{0.31\linewidth}
  \centering
  \includegraphics[width=2.7cm]{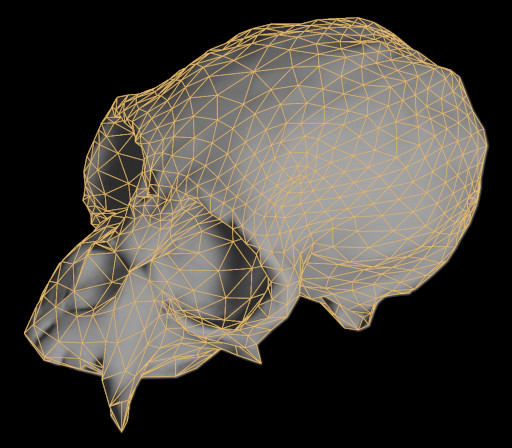}
  \subcaption{Nvdiffrast}
\end{minipage}
\begin{minipage}[t]{0.31\linewidth}
  \centering
  \includegraphics[width=2.7cm]{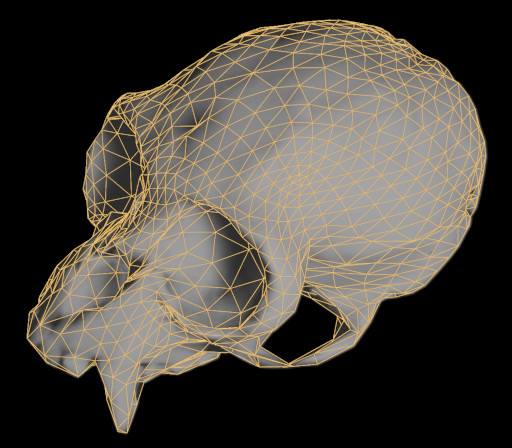}
  \subcaption{\textbf{Ours}}
\end{minipage}
\begin{minipage}[t]{0.31\linewidth}
  \centering
  \includegraphics[width=2.7cm]{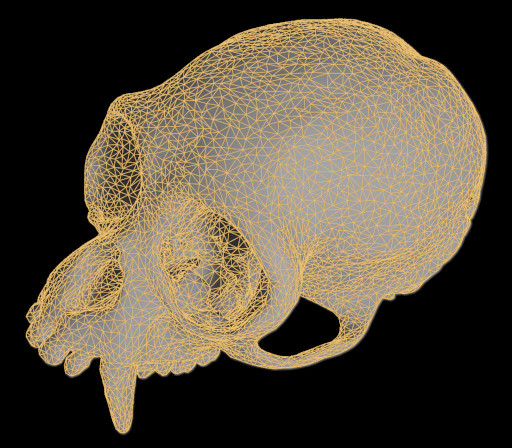}
  \subcaption{GT}
\end{minipage}
\caption{
  Visual comparison of optimized results with wire frame.
  (a) Nvdiffrast, (b) Ours, and (c) GT}
\label{fig:softras_cg_visual}
\EndFigure

%% file: supplements/stagepack_react_speed/table.tex
\begin{table}
    \centering
    \begin{tabular}{lcc}
        \multicolumn{1}{r}{Optimization target:} & Vertex position & Albedo texture  \\
        \hline \hline
        Ours full                                & \textbf{58.441} & \textbf{30.874} \\
        \quad-stage pack                         & 110.686         & 47.462          \\
        \quad-reactive cache                     & 317.814         & 306.594         \\
        \quad-stage pack, reactive cache         & 364.524         & 315.741         \\
        baseline                                 & 1780.90         & 1652.14         \\
        \hline
    \end{tabular}
    \caption{\label{tab:stage_packing_validation}
        Performance validation for stage packing and reactive cache.
        The average time (ms) of 1,000 iterations to optimize vertex positions and albedo texture on Arm Mali-G76.
        Avocado geometry is optimized with a single view.
        Ours full on the top is combined with stage pack, substage pack and reactive cache.
        Lower lines with "-" remove individual components.
        The baseline is naive shader implementation, and "-stage pack, reactive cache" (substage pack only) is a similar case to traditional kernel fusion technique.
    }
\end{table}

%% file: 5_experiments/experiments.tex
\section{Experiments}
\label{sec:experiments}

In this section, we evaluate our Dressi by comparing existing methods.
All experiments were run on a single NVIDIA GeForce RTX 2080, Intel(R) Core(TM) i7-9700K, and 32GB RAM.

\input{5_experiments/1_performance}
\input{5_experiments/2_softras}

%% file: 5_experiments/1_performance.tex
\input{supplements/speed_test/table}
\input{supplements/textured_speed_test/table}

\subsection{Performance Comparison}
\label{subsec:performance}
%
To validate that the performance of our renderer is
robust against several parameters,
we measured the computational costs to calculate forward and backward functions in two settings.
We compare Dressi with Nvdiffrast and PyTorch3D.
Because none of the methods supported the same shaders, we used shaders as similar as possible.
We show the average rendering time of 1,000 frames, removing outliers outside the $95\%$ confidence interval for all recorded data.

%
First, we evaluated the performance against the number of triangles and the resolution of window sizes.
To consider the effect of view-dependent geometry complexity, we rendered the meshes in six views with white vertex color and averaged the time spent on rendering.
Table~\ref{tab:non_textured_speed_test} shows
the performance of our method outperforms those of Nvdiffrast and PyTorch3D for all meshes and window resolutions.

%
Next, we demonstrate the scalability against the texture resolutions with shaders that map diffuse albedo textures \cite{cc0textures} to the meshes.
We changed the texture resolutions for each mesh with the fixed window resolution of $2048 \times 2048$.
Table~\ref{tab:textured_speed_test} shows that our performance is at least twice as fast as the others for all texture resolutions.

%% file: supplements/speed_test/table.tex
\begin{table}
  \centering
  \begin{tabular}{ccccc}
                             &             & \begin{minipage}{10mm}\centering\scalebox{0.04}{\includegraphics{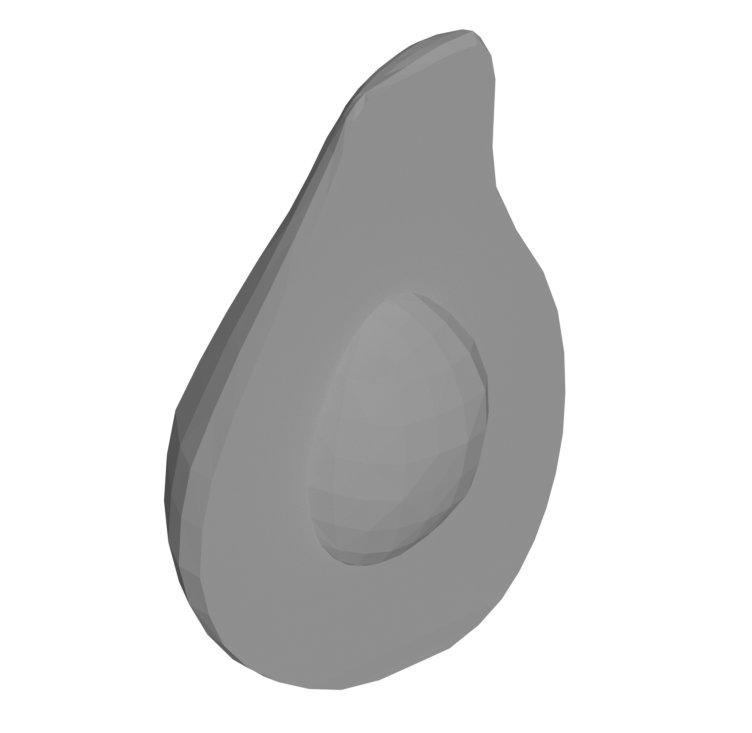}}\end{minipage}                         & \begin{minipage}{10mm}\centering\scalebox{0.04}{\includegraphics{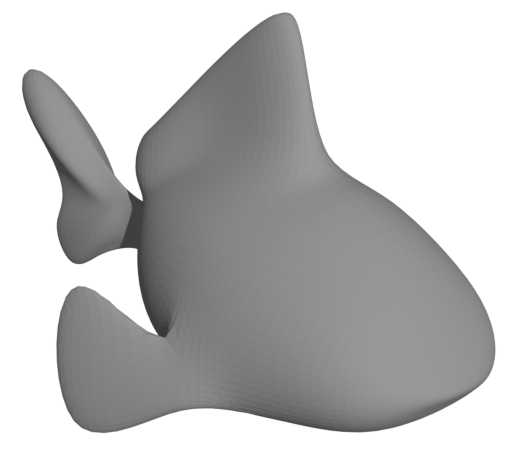}}\end{minipage} & \begin{minipage}{10mm}\centering\scalebox{0.04}{\includegraphics{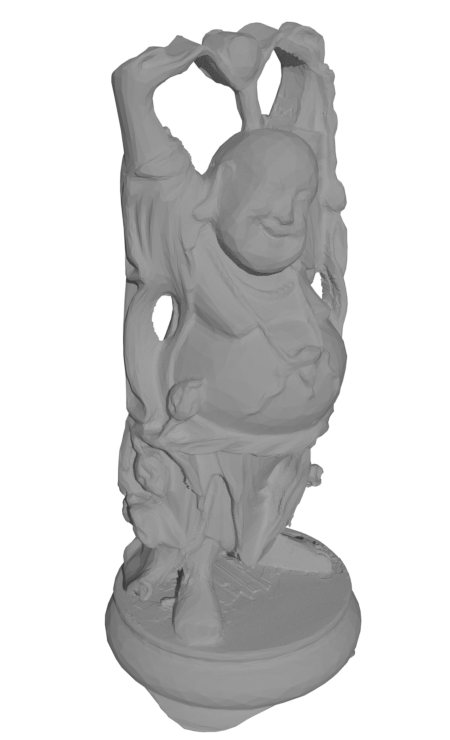}}\end{minipage} \\
                             & \#Vertices  & 406                                               & 7107                      & 25697                     \\
                             & \#Triangles & 682                                               & 14208                     & 51414                     \\
    \hline
                             & Window res. & \multicolumn{3}{c}{Fwd + bwd time per frame (ms)}                                                         \\
    \hline
                             & 256x256     & 0.304                                             & 0.308                     & 0.364                     \\
    \textbf{Ours}            & 512x512     & 0.442                                             & 0.480                     & 0.502                     \\
                             & 1024x1024   & 1.034                                             & 1.106                     & 1.104                     \\
                             & 2048x2048   & 3.301                                             & 3.545                     & 3.469                     \\
    \hline
                             & 256x256     & 1.060                                             & 0.480                     & 0.646                     \\
    Nvdiffrast               & 512x512     & 1.305                                             & 0.929                     & 0.853                     \\
    \cite{Laine2020diffrast} & 1024x1024   & 2.371                                             & 2.523                     & 2.086                     \\
                             & 2048x2048   & 5.403                                             & 8.010                     & 5.518                     \\
    \hline
                             & 256x256     & 8.054                                             & 10.272                    & 9.393                     \\
    PyTorch3D                & 512x512     & 12.570                                            & 15.354                    & 14.919                    \\
    \cite{ravi2020pytorch3d} & 1024x1024   & 32.136                                            & 39.034                    & 36.382                    \\
                             & 2048x2048   & 105.823                                           & 134.795                   & 121.624                   \\
    \hline
  \end{tabular}
  \caption{\label{tab:non_textured_speed_test}
    Performance comparison with non-textured meshes to optimize vertex positions.
    Ours and Nvdiffrast render silhouettes.
    For PyTorch3D, we apply Gouraud shading.
  }
\end{table}

%% file: supplements/textured_speed_test/table.tex
\begin{table}
  \centering
  \begin{tabular}{cccc}
                             &              & \begin{minipage}{15mm}\centering\scalebox{0.06}{\includegraphics{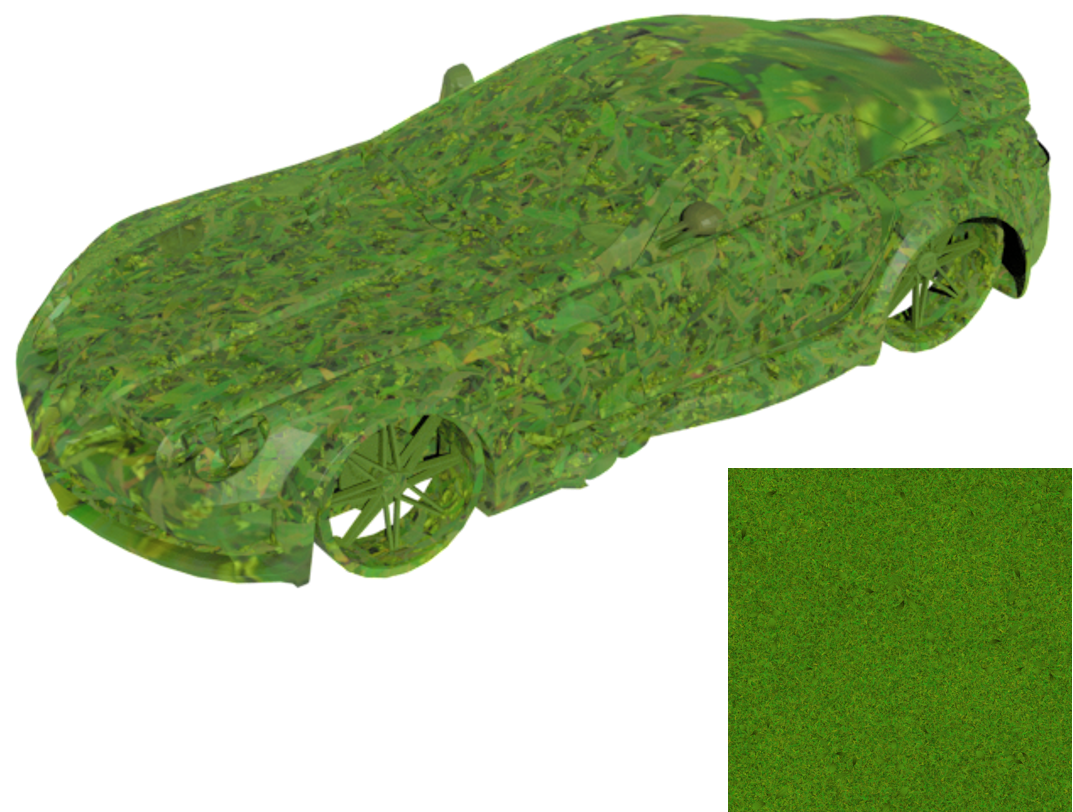}}\end{minipage}                         & \begin{minipage}{15mm}\centering\scalebox{0.06}{\includegraphics{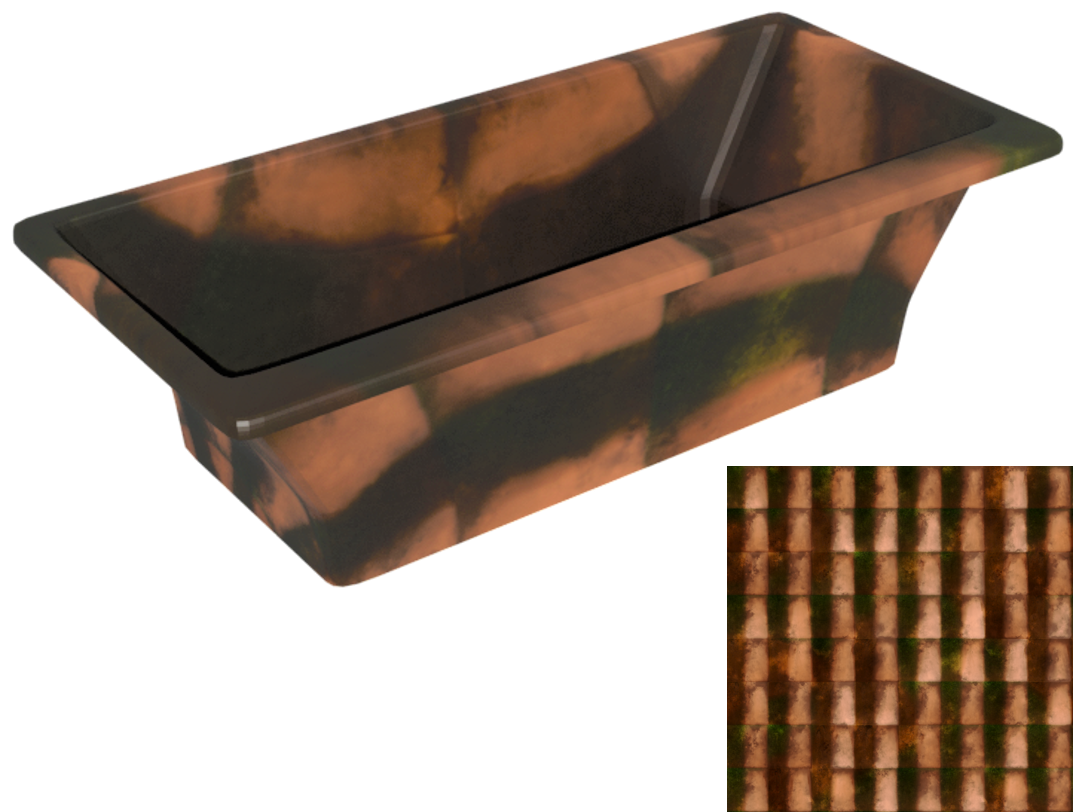}}\end{minipage} \\
                             & \#Vertices   & 31027                                             & 7232                      \\
                             & \#Triangles  & 55704                                             & 14217                     \\
    \hline
                             & Texture res. & \multicolumn{2}{l}{Fwd + bwd time per frame (ms)}                             \\
    \hline
                             & 1024x1024    & 2.929                                             & 2.893                     \\
    \textbf{Ours}            & 2048x2048    & 3.965                                             & 3.953                     \\
                             & 4096x4096    & 8.148                                             & 8.045                     \\
    \hline
                             & 1024x1024    & 6.584                                             & 6.254                     \\
    Nvdiffrast               & 2048x2048    & 8.922                                             & 8.653                     \\
    \cite{Laine2020diffrast} & 4096x4096    & 18.676                                            & 18.635                    \\
    \hline
                             & 1024x1024    & 137.681                                           & 118.184                   \\
    PyTorch3D                & 2048x2048    & 143.237                                           & 119.267                   \\
    \cite{ravi2020pytorch3d} & 4096x4096    & 141.667                                           & 120.017                   \\
    \hline
  \end{tabular}
  \caption{\label{tab:textured_speed_test}
    Performance comparison with textured-meshes to optimize diffuse albedo textures.
    Ours and Nvdiffrast render color images with unlit shading.
    We apply Phong shading for PyTorch3D.}
\end{table}

%% file: 5_experiments/2_softras.tex
%
\BeginFigure
\centering
\includegraphics[width=8.5cm]{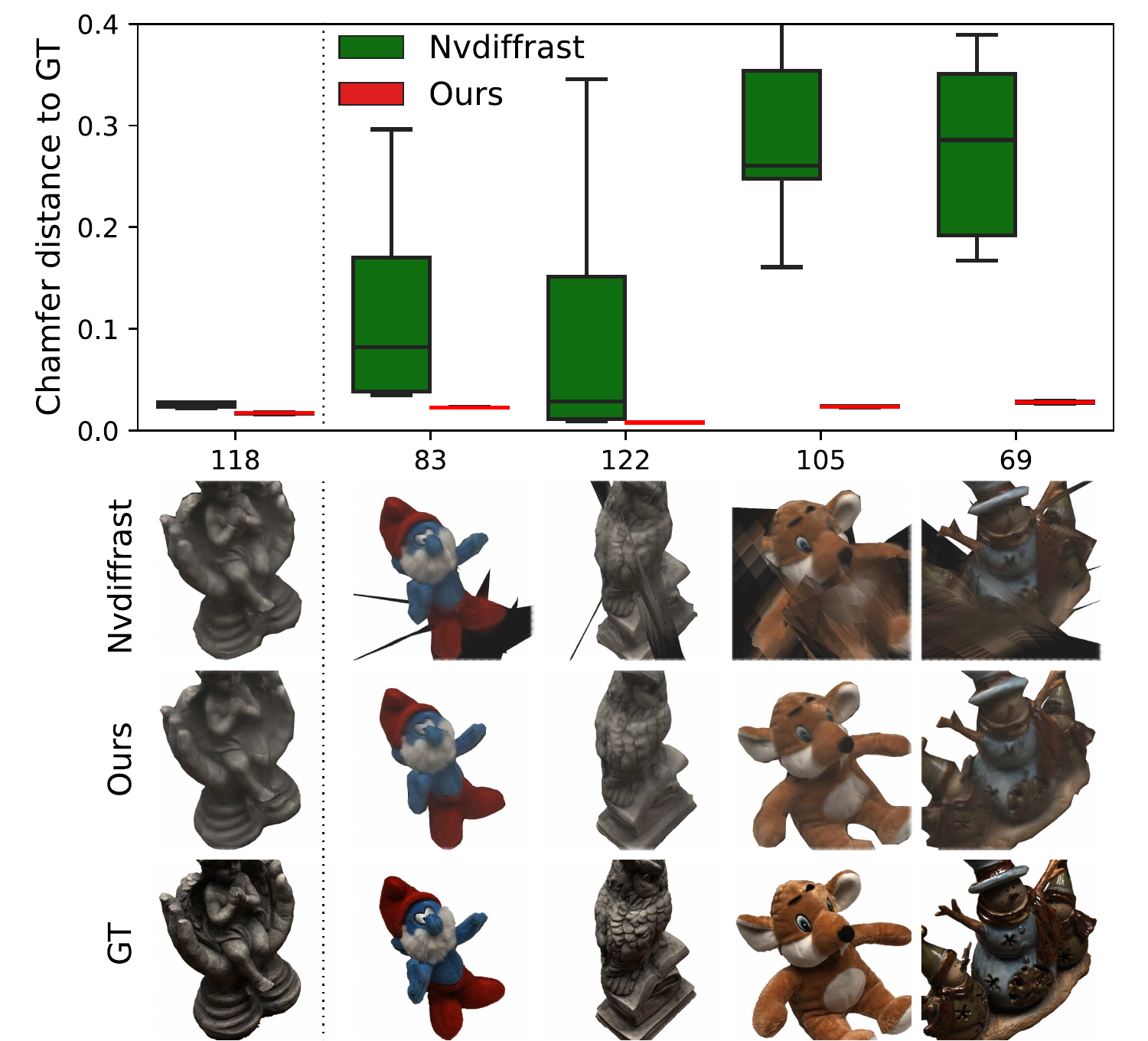}
\caption{
    The top row: box plot for 3D object reconstruction task with real data.
    Statistics with 10 trials of Chamfer distance are shown vertically, and the horizontal numbers are object IDs.
    Ours shows smaller variances and better distances for all objects.
    The bottom row: visual comparison with a reference GT image.
    We visualize optimized models corresponding to median Chamfer loss.
    Nvdiffrast and ours are rendered by GT camera parameters.
}
\label{fig:softras_real}
\EndFigure
%
%

%
\BeginFigure
\begin{minipage}[b]{0.35\linewidth}
    \centering
    \includegraphics[width=1.3cm]{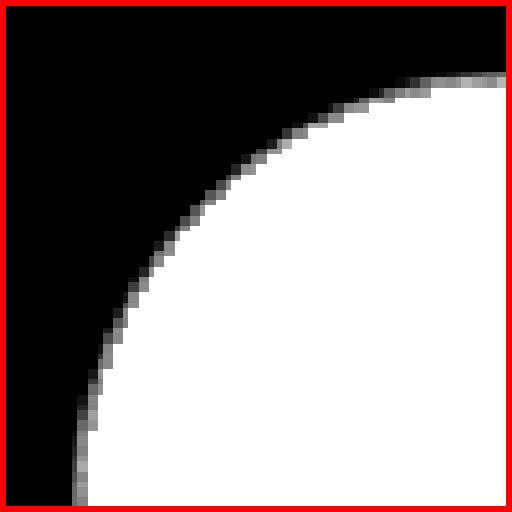}
    \subcaption{Ours (HardSoftRas)}
    \includegraphics[width=1.3cm]{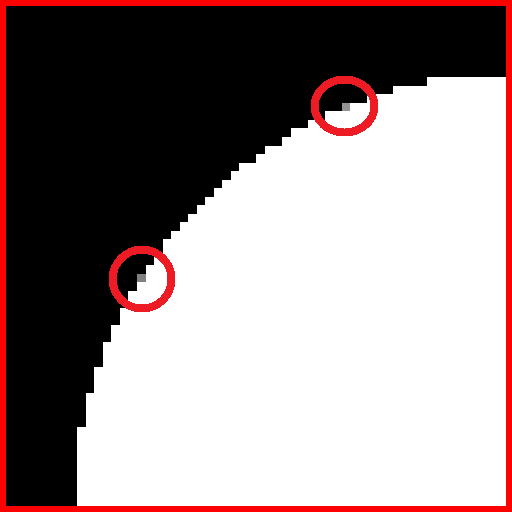}
    \subcaption{Nvdiffrast}
\end{minipage}
\begin{minipage}[b]{0.64\linewidth}
    \includegraphics[width=5.0cm]{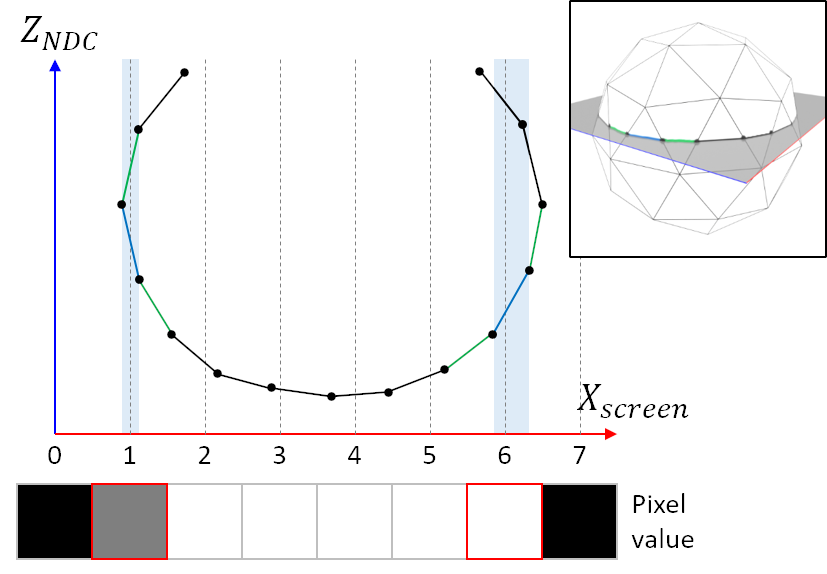}
    \subcaption{Geometric anti-aliasing of Nvdiffrast}
\end{minipage}
\caption{Sphere rendering comparison.
    (a) Our HardSoftRas blurs all edge pixels.
    (b) Nvdiffrast updates only two edge pixels in red circles.
    (c) How the geometric AA of Nvdiffrast works on a white curved surface with a black background.
    We show a slice of a simplified sphere in the upper right.
    Geometric edges are in continuous normalized device coordinate (NDC) space and are projected onto discrete pixels.
    Red pixels are target edge pixels, and the corresponding blue geometric edges are located on the pixel center and have the closest depth.
    See left red rectangle pixel affected by AA.
    Blue edge occludes a neighboring green edge at the pixel center.
    Thus, its color is blended into gray, and the gradient for the pixel is generated accordingly.
    In contrast, the pixel in the right red rectangle is not affected by AA because the corresponding blue edge does not occlude a neighboring green edge.
}
\label{fig:nv_analysis}
\EndFigure

\subsection{3D Reconstruction with Real Data}
\label{subsec:exp_real}
%
We also performed practical 3D reconstruction with Dressi and Nvdiffrast.
We used a real dataset for multi-view object reconstruction, DTU dataset \cite{aanaes2016large}, and its supplemental silhouettes \cite{yariv2020multiview}.
A big difference from the HardSoftRas validation is the limited number of views.
Only 49 or 64 views are available for each object.
Another challenge is the color discrepancy between the views caused by the lack of perfect control over the camera settings.
We fix the camera parameters as GT and minimize the L2 loss of color and silhouettes in $200 \times 150$ resolution with unlit shading using the Adam optimizer.
Vertex positions and diffuse albedo texture are jointly optimized.
We start the optimization with a white sphere, and a Laplacian regularizer with scheduling is used.

%
The top row of \figurename~\ref{fig:softras_real} is a box plot with five real objects for our method and Nvdiffrast with 10 trials to consider non-determinism of GPUs.
The chamfer distance (lower is better) was used as the evaluation metric in this experiment.
First, we tuned the hyperparameters (learning rate and regularization factors) for Nvdiffrast with an object called ``118".
Then other four objects were optimized with the same hyperparameters.
Our method used the common hyperparameters tuned for Nvdiffrast and $\mathbf{r}=2\mathrm{pix}, \mathbf{K}=1$ for all five objects.
During optimization, we sometimes observed sudden shape collapse for Nvdiffrast, which is reflected as a large variance in the box plot.
In contrast, ours never suffers such collapse with slight variance among the trials.

%
To analyze these behaviors, we show the initial white sphere rendered by each method and how the AA of Nvdiffrast works in \figurename~\ref{fig:nv_analysis}.
Although (a) ours shows uniformly blurred edges, (b) Nvdiffrast selectively blurs the edge pixels.
As shown in (c), Nvdiffrast follows conventional geometric anti-aliasing criteria to select a limited number of edge pixels for color blending and gradient generation.
Therefore, the gradients of Nvdiffrast in the screen space are sparse and discontinuous, and optimization using it is unstable.
Our HardSoftRas, which blends all edge pixels, ensures dense and smooth far-range gradients, achieving more robust optimization.

%
The bottom row of \figurename~\ref{fig:softras_real} shows a visual comparison of the reconstruction results with the median accuracy and GT.
The results of Nvdiffrast suffer from shape artifacts.
Our method can successfully reconstruct real objects with complex high-curvature surfaces.

%% file: 6_applications/applications.tex
\section{Applications}
\label{sec:app}

The proposed differentiable renderer, Dressi, has the potential to support many applications.
We demonstrate that Dressi can render a photorealistic facial model with complex shaders for human skin, eyeballs, and hair materials.
The supplemental material shows more applications: environment map optimization, normal map optimization and human face modeling with 3D morphable model.

\input{6_applications/1_lysa}

%% file: 6_applications/1_lysa.tex
%
\BeginFigure
\centering
\includegraphics[width=\linewidth]{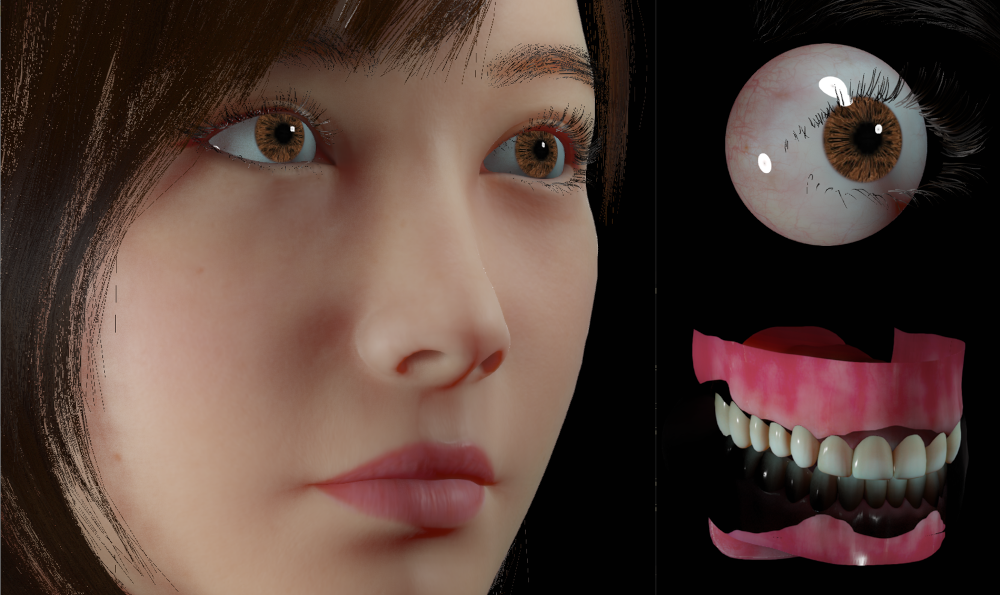}
\caption{High quality digital human rendered by Dressi in real-time.
    It contains skin, hair, eyeball, teeth shaders, and shadowing.}
\label{fig:lysa}
\EndFigure
%
%

%
\BeginFigure
\centering
\includegraphics[width=\linewidth]{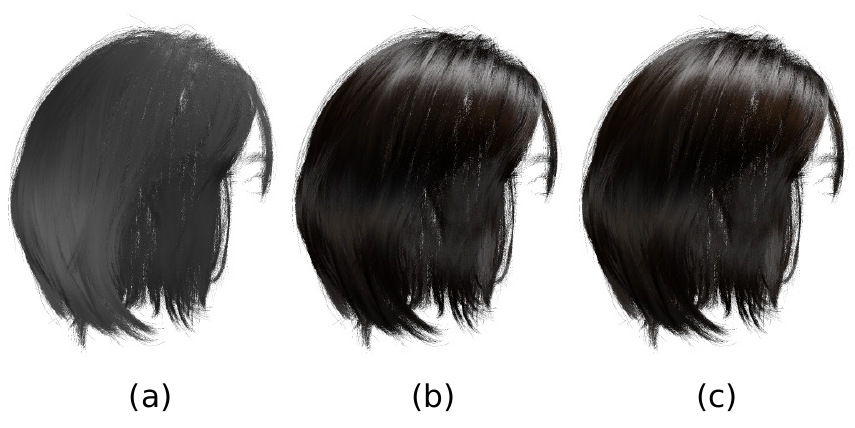}
\caption{
    Hair optimization:
    (a) rendered with initial parameter, (b) appearance after optimization of biophysical parameters, and (c) target image of the optimization.
    In the biophysical parameters, melanin components (melanin and melanin redness) affect hair color, and roughness parameters (radial and azimuthal roughness) control the highlight size, and IOR is for the highlight position.
    The parameters are shared over all strands.
    We optimized the parameters under the constraints of melanin and roughness from 0.0 to 1.0 and IOR in the range of from 1.0 to 2.0.
}
\label{fig:hair}
\EndFigure

\subsection{Practical Graphics Rendering with Complex Shaders}
\label{subsec:lysa}
%
We show that our system can support practical complex shaders, including hair, skin, eyeballs, and teeth shaders like recent game engines \cite{unreal_engine, unity}.
\figurename~\ref{fig:lysa} shows the rendering results of our high-quality digital human model with them.
Moreover, our engine supports the backward pass for the complex shaders.

%
To demonstrate inverse rendering with the complex shaders, we performed optimization of hair shading with the Marschner shader ~\cite{Marschner_2003_TOG}, which is a well-known physically-based reflection model for hair materials.
Biophysical parameters, including melanin, melanin redness, radial and azimuthal roughness, and index of refraction (IOR), were optimized.
The geometry and camera parameters were known, and an image from only one viewpoint was used.
\figurename~\ref{fig:hair} shows (a) the initial, (b) optimized, and (c) target images.
The optimized image (b) has the same appearance as the target image (c).

%% file: 7_conclusion/conclusion.tex
\section{Conclusion and Future Work}
%
In this paper, we proposed Dressi, a hardware-agnostic and highly efficient rasterization-based differentiable renderer to support various graphics hardware.
Dressi is based on a new design in which DR algorithms are completely written by AD for DR.
Our Dressi-AD realizes hardware independence by Vulkan API and an inverse UV technique.
Dressi-AD employs stage packing, a runtime optimization method with a reactive cache to accelerate computational graph executions and adapt hardware constraints.
HardSoftRas, our proposed rendering process entirely implemented on Dressi-AD, is the first DR algorithm to generate screen space far-range gradients under the limitation of graphics pipelines.
We demonstrated that our approach successfully works on a variety of graphics hardware (i.e., NVIDIA, AMD, Intel, and Arm).
We validated that our stage packing is adaptive to hardware constraints and contributes to speed.
Our Dressi has the potential to realize a wide range of practical applications in various devices.
Our experiments with synthetic and real data demonstrated that the speed and quality of our renderer outperform those of other state-of-the-art methods.
Our rendering for a digital human is of high quality and can run backward passes.
We show the optimization of hair color, which is not supported by other rasterization-based DR frameworks.

%
We leave two research topics for more practical problems in our framework.
The first topic is about the propagation of gradients of vertex attributes.
HardSoftRas employs depth peeling to consider gradients of triangles occluded in screen space.
However, the computational cost increases in proportion to many z-buffers.
Although we prioritize important triangles to reduce the number of z-buffers,
other order-independent transparency methods may realize more efficient gradient propagation.
%
The second topic is the further flexibility of our renderer.
Combinations of DR and neural networks are often used, such as optimization with identity loss \cite{gecer2019ganfit}.
We do not introduce operators of neural networks into our framework.
Connecting common AD libraries through the CUDA interface is possible, but this approach sacrifices hardware independence.
We can define the standard operators of neural networks in basic operations of GLSL functions,
and they provide more flexible solutions for practical problems.
%
We believe that Dressi contributes to the further improvement of rasterization-based DR, including the solutions of the referred two topics.

%% file: misc/acknowledgement.tex
\begin{acks}
    We thank the anonymous reviewers for their constructive comments, which have helped us improve the manuscript.
    We also want to thank Toshiya Hachisuka for helpful discussions.
    We are grateful to Naoya Iwamoto for designing visual materials.
    We appreciate Morales Espinoza Carlos Emanuel and Zhengqing Li for discussing related work and their thoughtful comments.
    We thank Naoya Hirai and Chun Geng for the appearance of skin shading.
    We are also grateful to Baris Gecer, Athanasios Vogiannou, Stylianos Moschoglou, Stylianos Ploumpis, Alexandros Lattas, and Stefanos Zafeiriou for the evaluation of GANFit \cite{gecer2019ganfit} and GANFit++ \cite{gecer2021fastganfit}.
\end{acks}

%% file: 9_appendix/appendix.tex


\input{9_appendix/1_code}

\section{Additional Applications}
\input{9_appendix/3_env}
\input{9_appendix/4_normal}
\input{9_appendix/5_faceswap}

%% file: 9_appendix/1_code.tex
%
\lstset{
  language=c++,
  basicstyle={\ttfamily \tiny},
  identifierstyle={},
  commentstyle={\itshape \color[rgb]{0.2,0.7,0.3}},
  keywordstyle={\color[rgb]{0,0,1}},
  morekeywords={nullptr},
  ndkeywordstyle={\color[rgb]{0.14,0.47,0.4}},
  morendkeywords={uint32_t, size_t, string, vector, function, map, tuple, priority_queue, tie, std,
      CpuImage, Variable, Variables, Function, Functions, VType, ImgSize, ShaderType, BwdFunc, DressiAD, DressiBasicRenderer,
      InitStatus, Optimizer, Stage, Stages, SubStage, SubStages,
      VkImage, VkImageUsage, VkCommandBuffer, VkRenderPass, VkPipeline},
  stringstyle={\color[rgb]{0.41,0.22,0.14}},
  frame={tb},
  breaklines=true,
  columns=fixed,
  basewidth=0.5em,
  lineskip=-0.5ex,
  showstringspaces=false
}
\newcommand{\CodeStrong}[1]{\underline{\textbf{#1}}}
\newcommand{\Code}[1]{\texttt{#1}}
\newcommand{\codename}{Listing}

\section{Implementation Details}
\label{appendix:code}
%
To clarify the implementation of our proposed system,
we describe class structures, API, and algorithm details in C++ code style.
Constant, reference, pointer, parallelization, and other less important parts are omitted for readability.

\subsection{Example of the Optimization on Dressi Renderer}
\label{appendix:code_main}
%
The following program is an example of the vertex positions and normal map optimization on Dressi renderer with HardSoftRas and physically-based shading.
Although Dressi code for optimization looks similar to PyTorch, Dressi is define-and-run.
At first, a renderer instance is declared (details in \secname ~\ref{appendix:renderer}) and scene files are loaded.
Second, a forward computational graph is built for rendering and loss calculation.
Third, the vertex positions and normal map are specified as optimization targets by the requires-gradient flag as in PyTorch.
After these setups, a main optimization process is executed using \Code{execStep()}.
Finally, the optimized results are saved.
\input{9_appendix/codes/main.tex}

%
\subsection{Example of the Dressi Renderer}
\label{appendix:renderer}
\Code{DressiBasicRenderer} is an example of rendering and optimization based on Dressi system design.
It has a \Code{DressiAD} instance, which is DR-specialized AD, and the renderer considers only the forward pass and optimizers (e.g., SGD).
The implementation details of \Code{DressiAD} are described in \secname ~\ref{appendix:dressi_ad}.
Scene data loaded to the CPU memory are parsed as \Code{CpuImage} structure, which is a simple buffer of a 2D image.
Non-image data (e.g., vertex positions) are converted into 2D image representation for efficient rendering.
After \Code{CpuImage}s are loaded, \Code{Variable} objects are created to represent a computational graph.
It is a data structure for inputs and outputs of functions (details in \secname ~\ref{appendix:func_var}).
Forward rendering and optimizer process are described by assembling operators and namespace \Code{F} functions (i.e., \Code{F::}) for the \Code{Variable} objects, constructing a computational graph.
The operators and \Code{F::} functions wrap \Code{Function} objects, which have a forward GLSL code and backward generator.
This implementation direction is similar to common AD libraries such as PyTorch and TensorFlow.
Examples of the \Code{F::} functions are described in \secname ~\ref{appendix:func_op}, and the implementation details of \Code{Function} are described in \secname ~\ref{appendix:func_var}.
\Code{\CodeStrong{BuildBasicRenderGraph}()} is an example of rendering functions using HardSoftRas (details in \secname ~\ref{appendix:render_graph}).
The loss for the rendered image and an optimizer algorithm are set to the \Code{DressiAD} instance.
\input{9_appendix/codes/renderer.tex}

\subsection{Example of the Rendering Algorithm Using HardSoftRas}
\label{appendix:render_graph}
%
\Code{\CodeStrong{BuildBasicRenderGraph}()} describes a rendering example using the HardSoftRas algorithm.
All our DR algorithms are written in AD, and there are no special backward declarations.
\Code{\CodeStrong{BuildRasterize}()}, which contains $\mathsf{Enlarge}$ and $\mathsf{Shift}$ operations and depth peeling, rasterizes triangles.
\Code{\CodeStrong{BuildBlend}()} takes the outputs of \Code{\CodeStrong{BuildRasterize}()} and shading parameters, and it performs blending operations for color and silhouette images.
\input{9_appendix/codes/render_graph.tex}

\subsection{Function and Variable Objects}
\label{appendix:func_var}
%
\Code{Function}s and \Code{Variable}s are the basic structures for representing the graph structure of a computational flow.
They have cross-references for constructing computational graphs.
\Code{Function} has a GLSL snippet for its forward process and a generator function for the backward pass.
\input{9_appendix/codes/func_var_basic.tex}

\subsection{Examples of the Namespace \Code{F} Functions}
\label{appendix:func_op}
%
Regarding Function and Variable objects, concrete operators/functions such as addition and multiplication are defined in namespace \Code{F}.
Similar to common AD libraries, both forward native code and backward generation are described in \Code{F::} functions.
In contrast to common AD libraries, DR-specific functions such as \Code{F::Rasterize()} and \Code{F::PeelDepth()} are also included in our AD as no-backward functions.
%
Each function declaration contains a GLSL snippet (e.g. \Code{\{y\}=\{x0\}+\{x1\};}).
\Code{\{x0\}} and \Code{\{x1\}} represent the first and second inputs of the function, and \Code{\{y\}} indicates an output.
The snippet is used to generate a fragment or compute shader code in \secname ~\ref{appendix:substage_pack}.
The input and output markers can be replaced by actual variable names.
%
Most functions can be written as shader snippets, but rasterization cannot be written as such.
Therefore, \Code{F::Rasterize} function is specially marked as \Code{ShaderType::RASTER}.
\input{9_appendix/codes/func_op.tex}

\subsection{Dressi-AD}
\label{appendix:dressi_ad}
%
The Dressi-AD class has interfaces to set loss variables and an optimizer (\Code{setLossVar()} and \Code{setOptimizer()}, respectively), transfer data between CPU and GPU (\Code{sendImg()} and \Code{recvImg()}, respectively), and execute optimization iteration (\Code{execStep()}).
\Code{execStep()} contains setups for the backward pass construction (\secname ~\ref{appendix:backward}), optimizer function construction, computational graph traversal, substage packing (\secname ~\ref{appendix:substage_pack}), stage packing (\secname ~\ref{appendix:stage_pack}), Vulkan object creation (\secname ~\ref{appendix:vulkan_obj}) from stage and substage graphs, and Vulkan command execution.
\input{9_appendix/codes/engine.tex}

\subsection{Backward Pass Construction}
\label{appendix:backward}
%
Backward pass construction is mostly similar to common AD libraries.
\Code{\CodeStrong{BuildBackward()}} backward traverses a forward computational graph from loss variables, sums up gradient variables by chain rule, and returns a pair of input and corresponding gradient variables.
\input{9_appendix/codes/engine_backward.tex}

\subsection{Substage Packing}
\label{appendix:substage_pack}
%
In substage packing, function objects are packed into substages, whereas clean and cached ones are skipped to reduce the computational cost.
``Clean'' indicates that the output Variable of a function is marked as clean.
``Cached'' indicates that the Variable is used as a substage I/O, and its data is stored on GPU as a result of previous iteration.
In our implementation, packing starts from the last output of the computational graph using a greedy algorithm.
The active substage being currently processed is iteratively updated.
At each iteration, the function that has more edges is packed into the active substage under Vulkan constraints in a greedy manner.
If there are no packable functions into the active substage, the substage is switched to a new one.
\input{9_appendix/codes/engine_substage.tex}

\subsection{Stage Packing}
\label{appendix:stage_pack}
%
Stage packing, which packs substages into stages, follows the same strategy as the substage packing.
\input{9_appendix/codes/engine_stage.tex}

\subsection{Vulkan Object Creation}
\label{appendix:vulkan_obj}
%
After stage packing, stages and substages are parsed to Vulkan objects.
The input and output variables of substages are allocated as \Code{VkImage} on GPU.
We use only \Code{VkImage} instead of \Code{VkBuffer} because a graphics pipeline is more efficient than a compute pipeline and only a fragment shader can have multiple outputs as images.
To simplify the implementation, non-image data such as vertex attributes are also treated as images.
Stages for rasterization and fragment shaders are parsed into \Code{VkRenderPass} objects, and substages are parsed into \Code{VkPipeline} to be a subpass in a \Code{VkRenderPass}.
For a compute shader, one stage should have only one substage because there is no hierarchical structure.
Therefore, the stage is parsed as one \Code{VkPipeline}.
\input{9_appendix/codes/engine_vulkan.tex}

%% file: 9_appendix/codes/main.tex
\begin{lstlisting}[mathescape]
int main(int argc, char *argv[]) {
  $\CodeStrong{OptimizeVertexPostionAndNormalTexture}$();
  return 0;
}

void $\CodeStrong{OptimizeVertexPostionAndNormalTexture}$() {
  // Create renderer instance
  DressiBasicRenderer renderer;
  // Load scene data on CPU
  renderer.loadScene("initial_scene.gltf", "target_img.png");

  // Build a computational graph
  uint32_t K = 2;         // The number of peeling for HardSoftRas
  float r = 0.01f;        // Radius parameter for HardSoftRas
  float sigma = r / 7.f;  // Blending parameter for HardSoftRas
  float delta = r;        // Silhouette edge width for HardSoftRas
  float lr = 0.01f;       // Learning rate for optimizer
  renderer.buildGraph(K, r, sigma, delta, lr);
  // Mark vertex positions and a normal texture as optimization targets
  renderer.setRequiresGrad("vtx_pos");
  renderer.setRequiresGrad("normal");

  // Optimization interations
  for (int iter = 0; iter < 1000; iter++) {
    renderer.execStep();
  }

  // Save optimized data as a GLTF file
  renderer.saveScene("optimized_scene.gltf");
}
\end{lstlisting}

%% file: 9_appendix/codes/renderer.tex
\begin{lstlisting}[mathescape]
// DressiBasicRenderer: An example of renderer class for Dressi system.
class DressiBasicRenderer {
public:
  void loadScene(std::string gltf_filename,
                 std::string target_img_filename) {
    // Load initial scene from a file and parse to CpuImage structures
    std::tie(m_img_map["vtx_pos"], m_img_map["vtx_uv"], m_img_map["faces"],
             m_img_map["world_mat"], m_img_map["view_mat"],
             m_img_map["prj_mat"], m_img_map["env_img"],
             m_img_map["albedo"], m_img_map["metallic"],
             m_img_map["roughness"], m_img_map["normal"],
             m_img_map["background"]) = LoadGltfAsCpuImages(gltf_filename);
    // Load target an image file and parse to a CpuImage structure
    m_img_map["target"] = LoadTargetImageAsCpuImage(target_img_filename);

    // Create top variables of a computational graph
    m_var_map["vtx_pos"] = {VEC3, m_img_map["vtx_pos"].getImgSize()};
    m_var_map["vtx_uv"] = {VEC2, m_img_map["vtx_uv"].getImgSize()};
    m_var_map["faces"] = {IVEC3, m_img_map["faces"].getImgSize()};
    m_var_map["model_mat"] = {MAT4, {1, 1}};
    m_var_map["view_mat"] = {MAT4, {1, 1}};
    m_var_map["prj_mat"] = {MAT4, {1, 1}};
    m_var_map["env_img"] = {VEC3, m_img_map["env_img"].getImgSize()};
    m_var_map["albedo"] = {VEC3, m_img_map["albedo"].getImgSize()};
    m_var_map["metallic"] = {FLOAT, m_img_map["metallic"].getImgSize()};
    m_var_map["roughness"] = {FLOAT, m_img_map["roughness"].getImgSize()};
    m_var_map["normal"] = {VEC3, m_img_map["normal"].getImgSize()};
    m_var_map["background"] = {VEC3, m_img_map["background"].getImgSize()};
    m_var_map["target"] = {VEC4, m_img_map["target"].getImgSize()};

    // Set a flag to send CPU images to GPU
    m_is_sent = false;
  }

  void buildGraph(uint32_t K, float r, float sigma, float delta,
                  float lr) {
    // Build a rendering computational graph
    Variable rendered_img = $\CodeStrong{BuildBasicRenderGraph}$(
            m_var_map["vtx_pos"], m_var_map["vtx_uv"], m_var_map["faces"],
            m_var_map["model_mat"], m_var_map["view_mat"],
            m_var_map["prj_mat"], m_var_map["env_img"],
            m_var_map["albedo"], m_var_map["metallic"],
            m_var_map["roughness"], m_var_map["normal"],
            m_var_map["background"], K, r, sigma, delta);
    // Take L1 loss
    Variable loss = F::Mean(F::Abs(m_var_map["target"] - rendered_img));

    // Set the loss and an optimizer to DressiAD
    m_dressi_ad.setLossVar(loss);
    m_dressi_ad.setOptimizer([=](Variables xs, Variables gxs) {
        // SGD for all inputs
        Variables updated_xs;
        for (size_t i = 0; i < xs.size(); i++) {
          updated_xs.push_back(xs[i] - gxs[i] * lr);
        }
        return updated_xs;
    });
  }

  void setRequiresGrad(std::string name) {
    // Set a requires-gradient flag
    m_var_map[name].setRequiresGradRecursively();
  }

  void execStep() {
    // Send CPU images to GPU if needed.
    if (!m_is_sent) {
      for (auto [name, var]: m_var_map) {
        m_dressi_ad.sendImg(var, m_img_map[name]);
      }
      m_is_sent = false;
    }

    // Execute one iteration of the optimization
    m_dressi_ad.execStep();
  }

  void saveScene(std::string gltf_filename) {
    // Receive all GPU images
    for (auto [name, var]: m_var_map) {
      m_img_map[name] = m_dressi_ad.recv(var);
    }
    // Save the optimized scene to a GLTF file.
    SaveGltfFromCpuImages(m_img_map);
  }

private:
  // DressiAD instance for DR-specialized AD
  DressiAD m_dressi_ad;
  // Scene data map for CpuImage
  std::map<std::string, CpuImage> m_img_map;
  // Scene data map for Variable corresponding to CpuImage
  std::map<std::string, Variable> m_var_map;
  // Internal flags
  bool m_is_sent = false;
};
\end{lstlisting}

%% file: 9_appendix/codes/render_graph.tex
\begin{lstlisting}[mathescape]
Variable $\CodeStrong{BuildBasicRenderGraph}$(Variable vtx_pos, Variable vtx_uv,
        Variable faces, Variables model_mat, Variable view_mat,
        Variable prj_mat, Variable env_img, Variable albedo,
        Variable metallic, Variable roughness, Variable normal,
        Variable background, uint32_t K, float r, float sigma,
        float delta) {
  // Irradiance map
  Variable irrad_img = BuildIrradianceSample(env_img);
  // Pre-filtered environment map for glossy material
  Variable pref_img = BuildPrefEnvironmentSample(env_img)
  // BRDF integration map
  Variable brdf_img = BuildBrdfIntegrationMap();

  /******** The beginning of HardSoftRas algorithm *******/
  // Rasterize and shade for each peeling plane
  auto prev_prj_depth = F::Float(0.f);
  Variables shaded_imgs, stencils, edge_dists;
  for (uint32_t plane_idx = 0; plane_idx < K; plane_idx++) {
    // Rasterize a single geometry
    auto [stencil, edge_dist, world_pos, world_nor, uv, prj_depth] =
              $\CodeStrong{BuildRasterize}$(vtx_pos, vtx_uv, faces, model_mat,
                             view_mat, prj_mat, prev_prj_depth, r);
    prev_prj_depth = prj_depth;   // Update peeling depth
    // Note: To rasterize multiple geometries, extra depth test is needed

    // Physically-based shading
    auto shaded_img = BuildPBS(stencil, world_pos, world_nor, uv,
                               albedo, metallic, roughness, normal,
                               irrad_img, pref_img, brdf_img);

    shaded_imgs.push_back(shaded_img);
    stencils.push_back(stencil);
    edge_dists.push_back(edge_dist);
  }

  // Blend shaded planes
  auto [blended_shaded_img, blended_silhouette_img] = $\CodeStrong{BuildBlend}$(
          shaded_imgs, stencils, edge_dists, background, K, sigma, delta);
  /********* The end of HardSoftRas algorithm ********/

  // Apply tone-map and gamma correction
  blended_shaded_img = BuildToneMap(blended_shaded_img);

  // Join silhouette image as alpha
  auto rendered_img = F::Vec4(blended_shaded_img, blended_silhouette_img);

  return rendered_img;
}

auto $\CodeStrong{BuildRasterize}$(Variable vtx_pos, Variable vtx_uv,
                    Variable faces, Variable model_mat, Variable view_mat,
                    Variable prj_mat, Variable prev_prj_depth, float r) {
  auto vtx_pos = F::Vec4(vtx_pos, 1.f);  // Cast to homogeneous coordinate
  vtx_pos = model_mat * vtx_pos;         // Apply model matrix

  // Compute vertex normals from positions
  auto vtx_nor = BuildNormalCompute(vtx_pos);
  // Apply view and projection matices
  auto vtx_prj = prj_mat * view_mat * vtx_pos;

  // Lookup the vertex buffer with faces to create per-triangle attributes
  auto tri_prj_0, tri_prj_1, tri_prj_2 = LookupFaces(vtx_prj, faces);
  auto tri_pos_0, tri_pos_1, tri_pos_2 = LookupFaces(vtx_pos, faces);
  auto tri_nor_0, tri_nor_1, tri_nor_2 = LookupFaces(vtx_nor, faces);
  auto tri_uv_0, tri_uv_1, tri_uv_2 = LookupFaces(vtx_uv, faces);

  // Enlarge(): Enlarge triangle vertex positions by scaling `r`
  // Notice that the number of triangles will change, and the original
  // vertex indices are stored.
  auto [large_prj_0, large_prj_1, large_prj_2, original_vtx_idxs] =
          BuildEnlarge(tri_prj_0, tri_prj_1, tri_prj_2, r);

  // Join triangle vertex positions to rasterize the single vertex buffer
  auto large_prj_flat = F::Stack(large_prj_0, large_prj_1, large_prj_2);
  // Rasterize enlarged triangles, and obtain screen space vertex indices
  auto screen_vtx_idxs = F::Rasterize(large_prj_flat, original_vtx_idxs);

  // SignedDist(): Compute pixel-to-edge distance inside enlarged triangles
  auto edge_dist = BuildEdgeDistance(tri_prj_0, tri_prj_1, tri_prj_2,
                                     screen_vtx_idxs);

  // Compute screen space barycentric coordinates
  auto bary_coords = BuildBaryCoord(tri_prj_0, tri_prj_1, tri_prj_2,
                                    screen_vtx_idxs);
  // Interpolate vertex attributes to be screen space
  auto prj_pos = BuildInterpolate(tri_prj_0, tri_prj_1, tri_prj_2,
                                  screen_vtx_idxs, bary_coords);
  auto world_pos = BuildInterpolate(tri_pos_0, tri_pos_1, tri_pos_2,
                                    screen_vtx_idxs, bary_coords);
  auto world_nor = BuildInterpolate(tri_nor_0, tri_nor_1, tri_nor_2,
                                    screen_vtx_idxs, bary_coords);
  auto uv = BuildInterpolate(tri_uv_0, tri_uv_1, tri_uv_2,
                             screen_vtx_idxs, bary_coords);

  // Shift(): Depth modification
  auto prj_depth = F::GetW(prj_pos);
  auto soft_depth = edge_dist * 0.5f + 0.5f;
  auto hard_depth = prj_depth * 0.5f;
  auto is_inside = (0.f < edge_dist);
  prj_depth = F::Mix(soft_depth, hard_depth, is_inside);
  prj_depth = F::SetFragDepth(prj_depth);  // Set depth to gl_FragDepth
  // Depth peeling
  auto stencil = F::PeelDepth(prj_depth, prev_prj_depth);

  return {stencil, edge_dist, world_pos, world_nor, uv, prj_depth};
}

auto $\CodeStrong{BuildBlend}$(Variables shaded_imgs, Variables stencils,
                Variables edge_dists, Variable background, uint32_t K,
                float sigma, float delta) {
  // Compute pixel weights from edge distances
  Variables weights;
  Variables weighted_cols;
  for (uint32_t plane_idx = 0; plane_idx < K; plane_idx++) {
    auto prob = 1.f / (1.f + F::Exp(edge_dists[i] / -sigma));
    auto weight = prob * stencils[plane_idx];
    weights.push_back(weight);
    weighted_cols.push_back(weight * shaded_imgs[plane_idx]);
  }

  // Blend shaded color by normalized weights
  auto sum_stencils = F::SumPixelWise(stencils);
  auto sum_weight = F::SumPixelWise(weights) / sum_stencils;
  auto sum_weighted_col = F::SumPixelWise(weighted_cols) / sum_stencils;
  auto blended_col = F::Mix(background, sum_weight, sum_weighted_col);

  // Blend silhouettes
  auto blended_sil = F::Float(1.f);
  for (uint32_t plane_idx = 0; plane_idx < K; plane_idx++) {
    blended_sil *= (1.f - weights[plane_idx]);
  }
  blended_sil = 1.f - blended_sil;

  // Blend hard faces by an edge mask
  auto is_hard = (0.f < edge_dists[0]);
  auto hard_shaded = shaded_imgs[0] * is_hard;  // Mask inside
  auto edge_mask = BuildEdgeMask(edge_dists[0], delta);
  blended_col = F::Mix(hard_shaded, blended_col, edge_mask);
  blended_sil = F::Mix(is_hard, blended_sil, edge_mask);

  return {blended_col, blended_sil};
}
\end{lstlisting}

%% file: 9_appendix/codes/func_var_basic.tex
\begin{lstlisting}[mathescape]
// Variable Data Types
enum VType { FLOAT, VEC2, ..., MAT2, ..., INT, IVEC2, ... };

// Variable Object
class Variable {
public:
  Variable(VType vtype = FLOAT, ImgSize img_size = {1, 1}):
      m_vtype(vtype), m_img_size(img_size) {}

  // Getters / setters
  VType getVType() { ... }
  void setVType(VType vtype) { ... }
  ImgSize getImgSize() { ... }
  void setImgSize(ImgSize img_size) { ... }
  Function getCreator() { ... }
  void setCreator(Function creator) { ... }
  std::vector<Function> getUsers() { ... }
  void addUser(Function user) { ... }
  bool getRequiresGrad() { ... }
  void setRequiresGrad(bool req_grad) { ... }
  void setRequiresGradRecursively(bool req_grad) { ... }
  bool IsDirty() { ... }
  void setDirty(bool is_dirty) { ... }
  void setDirtyRecursively(bool is_dirty) { ... }

private:
  VType m_vtype;
  ImgSize m_img_size;
  Function m_creator = Empty;
  std::vector<Function> m_users;
  bool m_req_grad = false;
  bool m_is_dirty = true;
};

// Aliases for simplification
using Variables = std::vector<Variable>;
using BwdFunc = std::function<Variable(Variables xs, Variable y,
                                       Variable gy, uint32_t bwd_idx)>;
// Shader type
enum ShaderType { FRAG, COMP, RASTER };

// Function Object
class Function {
public:
  Function(std::string code, ShaderType type, BwdFunc bwd_func) :
    m_fwd_code(code), m_type(type), m_bwd_func(bwd_func) {}

  // Getters
  std::string getFwdCode() { ... }
  ShaderType getShaderType() { ... }
  Variables getInputVars() { ... }
  Variable getOutputVar() { ... }

  // Building forward/backward connections
  Variable buildFwd(Variables xs) {
    m_xs = xs;
    for (auto x: xs) { x.addUser(this); }
    m_y.setCreator(this);
    // Infer and set variable type and image size of output
    m_y.setVType(InferOutputVType(xs));
    m_y.setImgSize(InferOutputImgSize(xs));
    return m_y;
  }
  Variable buildBwd(Variable gy, uint32_t bwd_idx) {
    // Create the graph for backward computation
    return m_bwd_func(m_xs, m_y, gy, bwd_idx);
  }

private:
  std::string m_fwd_code;
  ShaderType m_type;
  BwdFunc m_bwd_func;
  Variables m_xs;
  Variable m_y;
};

// Alias for simplification
using Functions = std::vector<Function>;
\end{lstlisting}

%% file: 9_appendix/codes/func_op.tex
\begin{lstlisting}[mathescape]
// Function Operators
namespace F {

Variable Float(float fv) {  // Constant float in GLSL
  return Function("{y}=float(" + fv + ");", FRAG, [](xs, y, gy, bwd_idx) {
    return nullptr;  // No backward
  }).buildFwd({});
}

Variable Add(Variable x0, Variable x1) {
  return Function("{y}={x0}+{x1};", FRAG, [](xs, y, gy, bwd_idx) {
    return gy;
  }).buildFwd({x0, x1});
}

Variable Mul(Variable x0, Variable x1) {
  return Function("{y}={x0}*{x1};", FRAG, [](xs, y, gy, bwd_idx) {
    if (bwd_idx == 0) {
      return gy * xs[1];  // Backward pass toward `x0`
    } else {
      return gy * xs[0];  // Backward pass toward `x1`
    }
  }).buildFwd({x0, x1});
}

Variable Sin(Variable x) {
  return Function("{y}=sin({x0});", FRAG, [](xs, y, gy, bwd_idx) {
    return F::Cos(gy);
  }).buildFwd({x});
}

Variable Rasterize(Variable vtx_pos, Variable vtx_attrib) {
  return Function("{y}={x1};",  // Rasterizing an attribute `vtx_attrib`
        RASTER,                 // Marked as rasterization specially
        [](xs, y, gy, bwd_idx) { return nullptr; }  // No backward pass
  ).buildFwd({x});
}

Variable PeelDepth(Variable frag_depth, Variable prev_frag_depth) {
  return Function("if({x0}<={x1})discard; {y}=1.0;", FRAG,
      [](xs, y, gy, bwd_idx) { return nullptr; }  // No backward pass
  ).buildFwd({frag_depth, prev_frag_depth});
}

...
}
\end{lstlisting}

%% file: 9_appendix/codes/engine.tex
\begin{lstlisting}[mathescape]
// DressiAD: DR-specialized AD library class
class DressiAD {
public:

  // Optimizer function
  //  (takes inputs and thier gradients, and returns optimized outputs.)
  using Optimizer = std::function<Variables(Variables xs, Variables gxs)>;

  // Setter
  void setLossVar(Variable loss_var) {
    m_loss_var = loss_var;
    m_init_status = InitStatus::BACKWARD; // Execute initialize process
  }
  void setOptimizer(Optimizer optim_func) { ... }

  // Building status
  enum InitStatus { BACKWARD, OPTIMIZER, TRAVERSE, SUBSTAGE, STAGE,
                    VULKAN, FINISHED};

  // Execute one step of rendering and optimization
  void execStep() {
    // Check dirty flags of input variables
    if (IsAnyVariableDirtyChanged(m_input_vars)) {
      m_graph_static_cnt = 0;
    }
    // Check rebuild condition
    if (m_graph_static_cnt == FAST_REBUILD_COUNT) {
      m_init_status = InitStatus::STAGE; // Execute fast rebuild
    } else if (m_graph_static_cnt == FULL_REBUILD_COUNT) {
      m_init_status = InitStatus::SUBSTAGE; // Execute full rebuild
    }

    if (m_init_status <= InitStatus::BACKWARD) {
      // 1) Traverse the forward computational graph and
      //    generate the backward graph by Function::buildBwd()
      m_input_vars, m_input_grad_vars = $\CodeStrong{BuildBackward}$(m_loss_var);
    }
    if (m_init_status <= InitStatus::OPTIMIZER) {
      // 2) Build the optimizer
      //    to connect the forward pass with the backward pass
      m_updated_vars = m_optim_func(m_input_vars, m_input_grad_vars);
      m_upd_inp_map = CreateMap(m_input_vars, m_updated_vars);
    }
    if (m_init_status <= InitStatus::TRAVERSE) {
      // 3) Traverse full computational graph
      //    from updated to input variables
      m_all_funcs = TraverseFuncs(m_updated_vars, m_input_vars);
    }
    if (m_init_status <= InitStatus::SUBSTAGE) {
      // 4) Pack the computational graph into SubStages
      m_substages = $\CodeStrong{PackDirtyFuncsIntoSubStages}$(m_all_funcs, m_vk_imgs);
    }
    if (m_init_status <= InitStatus::STAGE) {
      // 5) Pack the SubStages into Stages
      m_stages = $\CodeStrong{PackDirtySubStagesIntoStages}$(m_substages, m_vk_imgs);
    }
    if (m_init_status <= InitStatus::VULKAN) {
      // 6) Parse to Vulkan objects
      m_vk_imgs, m_vk_cmd_buf, m_vk_renderpasses, m_vk_pipelines, ... =
          $\CodeStrong{ParseStagesAsVulkanObjects}$(m_stages, m_vk_imgs, m_upd_inp_map);
    }

    // 7) Execute a command buffer
    VkQueueSubmit(m_vk_cmd_buf);

    // Mark as clean recursively
    for (auto v: m_input_vars) { v.setDirtyRecursively(false); }
    m_init_status = FINISHED;
  }

  // Transfer image data between CPU and GPU.
  void sendImg(Variable var, CpuImage cpu_img) {
    if (m_vk_imgs.contains(var)) {
      var.setDirty(true);  // Mark as changed
      SendHostImageToDevice(m_vk_imgs[var], cpu_img);  // CPU -> GPU
    }
  }
  CpuImage recvImg(Variable var) {
    if (m_vk_imgs.contains(var)) {
      return ReceiveHostImageFromDevice(m_vk_imgs[var]);  // GPU -> CPU
    }
    return nullptr;
  }

private:
  InitStatus m_init_status = InitStatus::BACKWARD;
  Variable m_loss_var;   // Last variable of the forward pass
  Optimizer m_optim_func;
  uint32_t m_graph_static_cnt = 0;

  Variables m_input_vars;      // Top variables of the forward pass
  Variables m_input_grad_vars; // Last variables of the backward pass
  Variables m_updated_vars;    // Last variables of the computational graph
  Functions m_all_funcs;       // All functions of the computational graph
  std::map<Variable, Variable> m_upd_inp_map;  // Map from input to updated

  SubStages m_substages;
  Stages m_stages;

  std::map<Variable, VkImage> m_vk_imgs;
  VkCommandBuffer m_vk_cmd_buf;
  std::vector<VkRenderPass> m_vk_renderpasses;
  std::vector<VkPipeline> m_vk_pipelines;
  ... // Many Vulkan objects
};
\end{lstlisting}

%% file: 9_appendix/codes/engine_backward.tex
\begin{lstlisting}[mathescape]
std::tuple<Variables, Variables> $\CodeStrong{BuildBackward}$(Variable loss_var) {
  // Mapping from a forward variable to backward gradient variables
  std::map<Variable, Variables> fwd_bwds_map;
  // Function queue that keeps the order of use
  std::priority_queue<Function> func_queue;

  // Set loss as the starting point
  fwd_bwds_map[loss_var] = F::Float(1.f);
  func_queue.push(loss_var.getCreator());

  // Traversal loop
  while (!func_queue.empty()) {
    // Pop the latest-used function.
    Function func = func_queue.top();
    func_queue.pop()
    if (IsSeenFunc(func)) continue;

    // Sum up gradients.
    Variable y = func.getOutputVar();
    Variables gys = fwd_bwds_map.at(y);
    Variable gy = F::SumPixelWise(gys);
    fwd_bwds_map.erase(y);

    Variables xs = func.getInputVars();
    for (size_t x_idx = 0; x_idx < xs.size(); x_idx++) {
      auto x = xs[x_idx];
      if (!x.getRequiresGrad()) continue;  // Skip no gradient path

      // Build a backward connection
      Variable gx = func.buildBwd(gy, x_idx);
      if (!gy) {
        continue;  // Skip non-backwardable function
      }
      // Register a new gradient
      fwd_bwds_map[x].push_back(gx);

      // Push a creator function for recurrent traversal
      Function x_creator = x.getCreator();
      if (x_creator) func_queue.push(x_creator);
    }
  }

  // Collect input and gradient of input variables
  Variables input_vars, input_grad_vars;
  for (auto [x, gxs]: fwd_bwds_map) {
    input_vars.push_back(x);
    input_grad_vars.push_back(F::SumPixelWise(gxs));
  }
  return {input_vars, input_grad_vars};
}
\end{lstlisting}

%% file: 9_appendix/codes/engine_substage.tex
\begin{lstlisting}[mathescape]
struct SubStage {
  Variables vtx_vars; // Vertex buffer inputs for rasterization
  Variables inp_vars; // Input attachment
  Variables tex_vars; // Texture sampler inputs
  Variables slt_vars; // Sampler-less texture inputs
  Variables uif_vars; // Uniform inputs
  Variables out_vars; // Color attachment (outputs)
  Variables gen_vars; // All generated variables including shader inside
  Functions funcs;
  std::string shader_code;
};
using SubStages = std::vector<SubStage>;

SubStages $\CodeStrong{PackDirtyFuncsIntoSubStages}$(
      Functions all_funcs, std::map<Variable, VkImage> cached_imgs) {
  // Traverse functions, ignoring clean and cached branches.
  auto dirty_funcs = RemoveCleanFuncs(all_funcs, cached_imgs);

  // Graph optimization
  OmitConstantFuncs(dirty_funcs);   // Precompute constant values
  OmitDuplicatedFuncs(dirty_funcs); // Omit same functions with same inputs

  // Search suitable packing under Vulkan limitations
  SubStages substages = $\CodeStrong{SearchSuitableFunctionPacking}$(dirty_funcs);

  for (auto substage: substages) {
    // Generate GLSL shader codes by string manipulation as following.
    //  1) Collect GLSL snippets of functions in a substage.
    //      ex.) "{y}={x0}+{x1};", "{y}=sin({x0});"
    //  2) Join snippet lines, and replace input and output markers.
    //      ex.) "v7=v5+v6; v8=sin(v7);"
    //  3) Add variable declaration, I/O codes of input attachments,
    //     main function, and description of attachments/uniforms/etc...
    //      ex.) "layout(...) uniform subpassInput sub_inp[2]; ...
    //            void main() { float v5=subpassLoad(sub_inp[0]); ...
    //                          float v7=v5+v6; float v8=sin(v7); ... }"
    substage.shader_code = GenerateShaderCode(substage);
  }
  return substages;
}

SubStages $\CodeStrong{SearchSuitableFunctionPacking}$(Functions dirty_funcs) {
  SubStages substages;
  SubStage active_substage;
  Variables used_vars;

  while (!dirty_funcs.empty()) {
    // Collect functions whose outputs are not used in other functions.
    Functions candidate_funcs = CollectLatestFuncs(dirty_funcs);
    // Sort functions by edge numbers to the substage.
    // We assume that the number of edges correlates a probability to be
    // a better choice to maximize the size of substage.
    candidate_funcs = SortByEdgeNumbers(candidate_funcs, active_substage);

    // Try packing from the last of computational graph.
    bool is_found = false;
    for (auto func: candidate_funcs) {
      // Try push a function into the substage
      auto trial_substage = $\CodeStrong{PushFrontFuncIntoSubStage}$(
              func, active_substage, used_vars);
      if ($\CodeStrong{IsSubStageVkLimitsSatisfied}$(trial_substage)) {
        active_substage = trial_substage;  // Suitable packing found
        dirty_funcs.erase(func);
        is_found = true;
        break;
      }
    }
    if (!is_found) {
      // Switch to a next substage if no more packing
      substages.push_back(active_substage);
      active_substage = {};  // Clear
      // Mask input variables as used for substage dependency
      for (auto inp_var: CollectAllInputVars(substage)) {
        used_vars.push_back(inp_var);
      }
    }
  }
  return substages;
}

SubStage $\CodeStrong{PushFrontFuncIntoSubStage}$(Function func, SubStage substage,
                                   Variables used_vars) {
  // Register function inputs
  for (auto inp_var: func.getInputVars()) {
    if (func.getShaderType() == RASTER) {
      substage.vtx_vars.push_back(inp_var);  // As vertex buffer
    } else if (IsSamplerType(inp_var)) {
      substage.tex_vars.push_back(inp_var);  // As texture sampler
    } else if (IsSamplerLessType(inp_var)) {
      substage.slt_vars.push_back(inp_var);  // As sampler-less texture
    } else if (inp_var.getImgSize() == {1, 1}) {
      substage.uif_vars.push_back(inp_var);  // As uniform
    } else {
      substage.inp_vars.push_back(inp_var);  // As input attachment
    }
  }

  // Remove generated variables from inputs (Vertex, texture,
  // and sampler-less texture must not be generated in the same substage.)
  Variables out_var = func.getOutputVar();
  substage.inp_vars.erase(out_var);
  substage.uif_vars.erase(out_var);

  // Register function output which is needed by other substages
  if (used_vars.contains(out_var)) {
    substage.out_vars.push_back(out_var);
  }
  // Register function output as generated variables
  substage.gen_vars.push_back(out_var);
  // Register function
  substage.funcs.push_back(func);
  return substage;
}

bool $\CodeStrong{IsSubStageVkLimitsSatisfied}$(SubStage substage) {
  // All output images must have same image size.
  if (!AreSameImgSizes(substage.out_vars)) return false;
  // All functions must have same shader type except for top rasterization.
  if (!AreSameShaderTypes(substage.funcs) &&
      !(substage.funcs[0].getShaderType() == RASTER &&
        AreSameShaderTypes(RemoveFirst(substage.funcs)))) return false;
  // Vertex/texture input must come from another substage.
  if (substage.gen_vars.containsAny(substage.vtx_vars)) return false;
  if (substage.gen_vars.containsAny(substage.tex_vars)) return false;
  // Limited numbers of Vulkan I/O
  if (MAX_VULKAN_INPUT_ATTACH < substage.inp_vars.size()) return false;
  if (MAX_VULKAN_TEXTURE_SAMPLER < substage.tex_vars.size()) return false;
  if (MAX_VULKAN_SAMPLED_IMAGE < substage.slt_vars.size()) return false;
  if (MAX_VULKAN_UNIFORM < substage.uif_vars.size()) return false;
  if (MAX_VULKAN_OUTPUT_ATTACH < substage.out_vars.size()) return false;
  ...  // Other number limitations for combined conditions
  return true;
}
\end{lstlisting}

%% file: 9_appendix/codes/engine_stage.tex
\begin{lstlisting}[mathescape]
struct Stage
{
  // I/O variables as same as SubStages'
  Variables vtx_vars, inp_vars, tex_vars, slt_vars, uif_vars, out_vars;
  // Hierarchical structure for SubStages
  SubStages substages;
};
using Stages = std::vector<Stage>;

Stages $\CodeStrong{PackDirtySubStagesIntoStages} $(
    SubStages all_substages, std::map<Variable, VkImage> cached_imgs)
{
  // Traverse a substage graph, ignoring clean and cached branches.
  auto dirty_substages = RemoveCleanSubStages(all_substages, cached_imgs);

  // Search suitable packing under Vulkan limitations.
  //   Packing strategy is same as `SearchSuitableFunctionPacking()`.
  Stages stages = SearchSuitableSubStagePacking(dirty_substages);
  return stages;
}
\end{lstlisting}

%% file: 9_appendix/codes/engine_vulkan.tex
\begin{lstlisting}[mathescape]
auto $\CodeStrong{ParseStagesAsVulkanObjects}$(Stages stages,
                                std::map<Variable, VkImage> prev_vk_imgs,
                                std::map<Variable, Variable> upd_inp_map) {
  // Collect image usages for substage I/O
  std::map<Variable, VkImageUsage> usages = CollectVkImageUsage(stages);
  // Create images with usages
  std::map<Variable, VkImage> vk_imgs;
  for (auto [var, usage]: usages) {
    if (prev_vk_imgs.contains(var)) {
      // Skip image creation to keep previous image data
      vk_imgs[var] = prev_vk_imgs[var];
    } else if (upd_inp_map.contains(var)) {
      // Use the same image for input and updated variables.
      // Inputs will be overwritten by updated ones for each iteration.
      vk_imgs[var] = upd_inp_map(var);
    } else {
      // Create a new Vulkan image on GPU
      vk_imgs[var] =
          VkCreateImage(var.getVType(), var.getImgSize(), usage, ...);
    }
  }

  // Create pipelines, render passes, and lots of other Vulkan objects
  VkCommandBuffer vk_cmd_buf;
  std::vector<VkRenderPass> vk_renderpasses;
  std::vector<VkPipeline> vk_pipelines;
  ...
  for (auto stage: stages) {
    // Build a graphics or compute pipeline according to stage type.
    ShaderType stage_type = GetShaderType(stage);
    if (stage_type == FRAG) { // Including `RASTER` too. Rasterization and
                              // fragment functions were packed into one
                              // stage.
      // Create a render pass from a stage for graphics pipelines.
      auto vk_renderpass = CreateRenderPass(stage);
      ...  // Many Vulkan setups
      VkCmdBeginRenderPass(vk_cmd_buf, vk_renderpass);

      // Subpass creation and recording
      for (auto substage: stage.substages) {
        // Create a pipeline for a substage as a subpass in the renderpass.
        auto vk_pipeline = CreateGraphicsPipeline(substage, vk_renderpass);
        vk_pipelines.push_back(vk_pipeline);
        ...  // Many Vulkan setups
        VkCmdBindPipeline(vk_cmd_buf, vk_pipeline);

        // Record drawing call
        if (substage.vtx_vars.empty()) {
          // Drawing for only fragment functions.
          //  No vertex buffer are bound, and dummy vertex shader is
          //  attached to rasterize a full-screen rectangle.
          VkCmdDraw(vk_cmd_buf);
        } else {
          // Drawing for rasterization and fragment functions.
          //  In order to rasterize one vertex attribute, a vertex buffer
          //  is bound, and a pass-through vertex shader is attached.
          VkCmdBindVertexBuffers(vk_cmd_buf);  // Bind vertex buffer
          VkCmdDraw(vk_cmd_buf);
        }
        VkCmdNextSubPass(vk_cmd_buf);
      }

      VkCmdEndRenderPass(vk_cmd_buf);
      vk_renderpasses.push_back(vk_renderpass);
    } else if (stage_type == COMP) {
      // Create a compute pipeline. It is slower than graphics pipelines.
      // `substages.size() == 1` because of no subpass for compute shaders.
      auto vk_pipeline = CreateComputePipeline(stage.substages[0]);
      vk_pipelines.push_back(vk_pipeline);
      ...  // Many Vulkan setups
      VkCmdBindPipeline(vk_cmd_buf, vk_pipeline);
      // Record drawing call
      VkCmdDispatch(vk_cmd_buf);
    }
  }
  return vk_imgs, vk_cmd_buf, vk_renderpasses, vk_pipelines, ...;
}
\end{lstlisting}

%% file: 9_appendix/3_env.tex
\subsection{Environment Map Optimization}
\label{subsec:env_opt}

To demonstrate the versatility of our renderer, we jointly optimize an environment map and the physically-based shading (PBS) \cite{mcauley2013PBR} roughness property.
We use three meshes whose materials have different metallic and roughness properties and the environment maps provided by a public repository \cite{hdrihaven}.
Fixed 20 views for 1,000 iterations are used to fit the pre-generated images rendered with GTs.
We use Adam \cite{kingma2014method} optimizer in the optimization process.
The environment map is initialized to a uniform gray color, and the minimum and maximum pixel intensities of the GT environment map are clamped to avoid noisy results.
\figurename~\ref{fig:env_opt} shows that every combination of the mesh and the environment map converges to GT value.
We observe some limitations through this experiment.
First, we cannot fit optimized values to ground truth values such as low metallic properties, high roughness properties, or highly complicated geometries.
It is impossible to optimize the environment map correctly with the matte materials due to the lack of visual cues in the rendered images.
Second, the optimization contains artifacts that are caused by sampling patterns.
There is room for improvement in the environment sampling method for optimization.

\BeginFigure
\centering
\includegraphics[width=\linewidth]{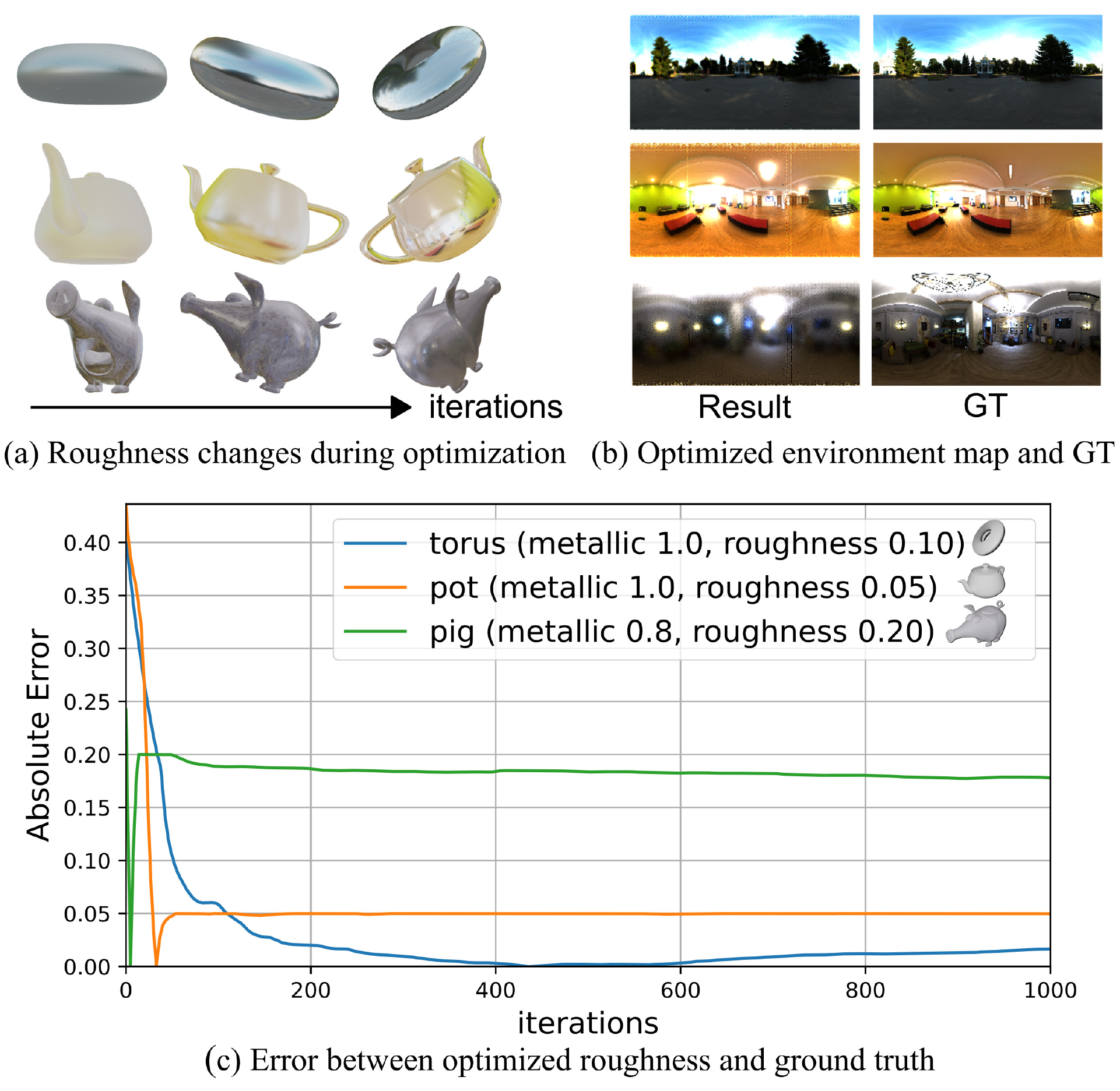}
\caption{
    Joint optimization of the PBS roughness property and the environment map: (a) changes in roughness through optimization, (b) optimized results and GTs of the environment maps, and (c) error between optimized roughness and GT for the meshes that are torus, pot, and pig \cite{keenan}.}
\label{fig:env_opt}
\EndFigure

%% file: 9_appendix/4_normal.tex
\subsection{Normal Map Optimization}
\label{subsec:normal_opt}

\BeginFigure
\centering
\includegraphics[width=\linewidth]{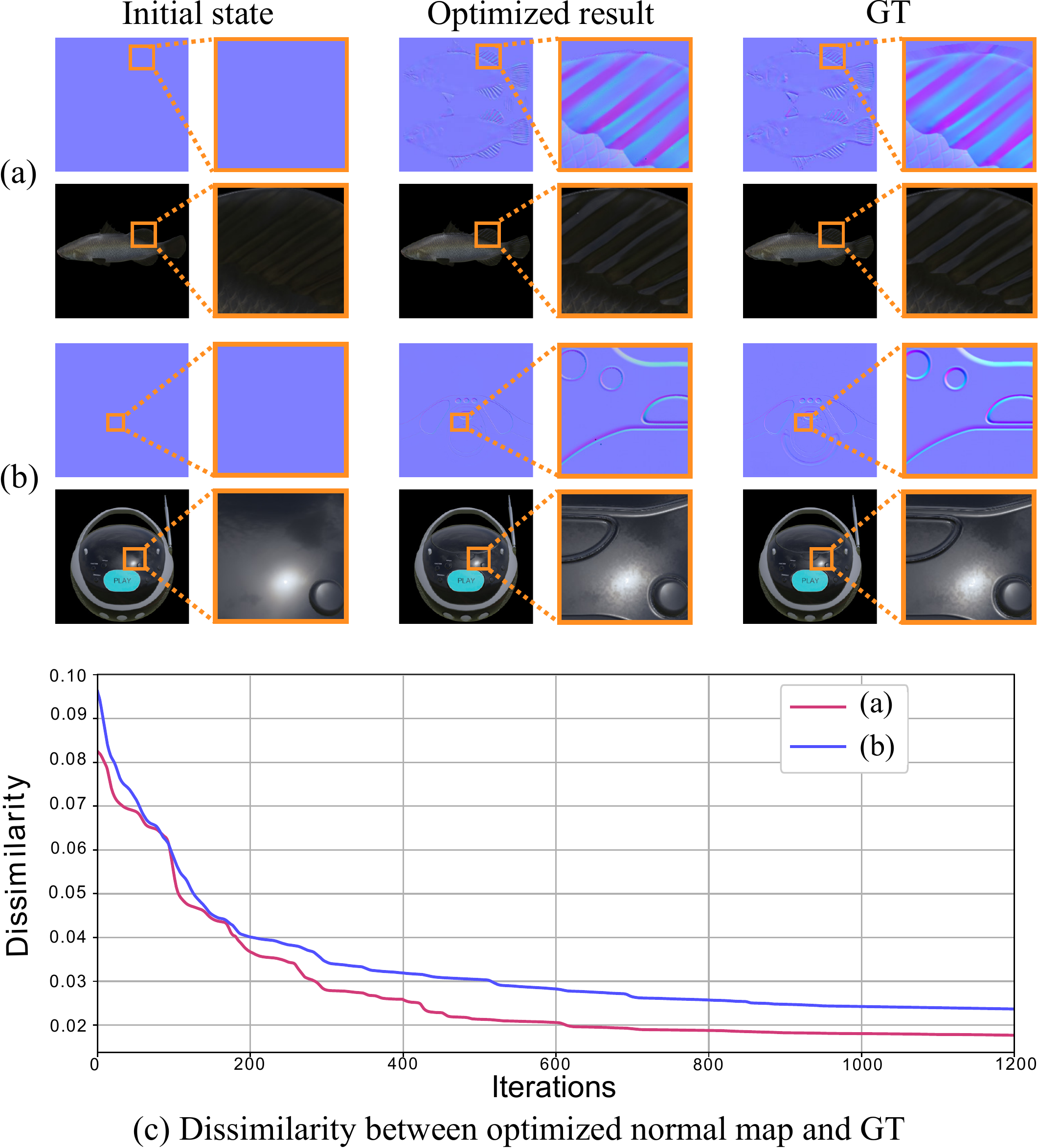}
\caption{
    Normal map optimization with fish and boombox models.
    The upper part shows the normal maps and rendered images of (a) fish and (b) boombox.
    The lower part (c) shows the numerical evaluation with cosine dissimilarity.
    The optimization starts from a uniform color image and finally generates a similar pattern with the ground truth after 1200 iterations.
    The cosine dissimilarity decreases with each iteration. This indicates that the optimized normal vector turns to be very close to the ground truth.}
\label{fig:normal_optimization}
\EndFigure

\BeginFigureTwoCol
\centering
\begin{minipage}[t]{0.07\linewidth}
    \centering
    {\!\!\!\!\!\!\!\!\!\!\!\!\!Input}
\end{minipage}
\begin{minipage}[t]{0.17\linewidth}
    \centering
    { \begin{tabular}{c}{GANFit}\\{\small \cite{gecer2019ganfit}} \end{tabular} }
\end{minipage}
\begin{minipage}[t]{0.18\linewidth}
    \centering
    { \begin{tabular}{c}{GANFit++}\\{\small \cite{gecer2021fastganfit}} \end{tabular} }
\end{minipage}
\begin{minipage}[t]{0.165\linewidth}
    \centering
    { \begin{tabular}{c}{hifi3dface}\\{\small \cite{hifi3dface2021tencentailab}} \end{tabular} }
\end{minipage}
\begin{minipage}[t]{0.155\linewidth}
    \centering
    { \begin{tabular}{c}{COLMAP}\\{\small \cite{schoenberger2016sfm,schoenberger2016mvs}}\end{tabular} }
\end{minipage}
\begin{minipage}[t]{0.16\linewidth}
    \centering
    {  {\textbf Ours} }
\end{minipage}

\includegraphics[width=\linewidth]{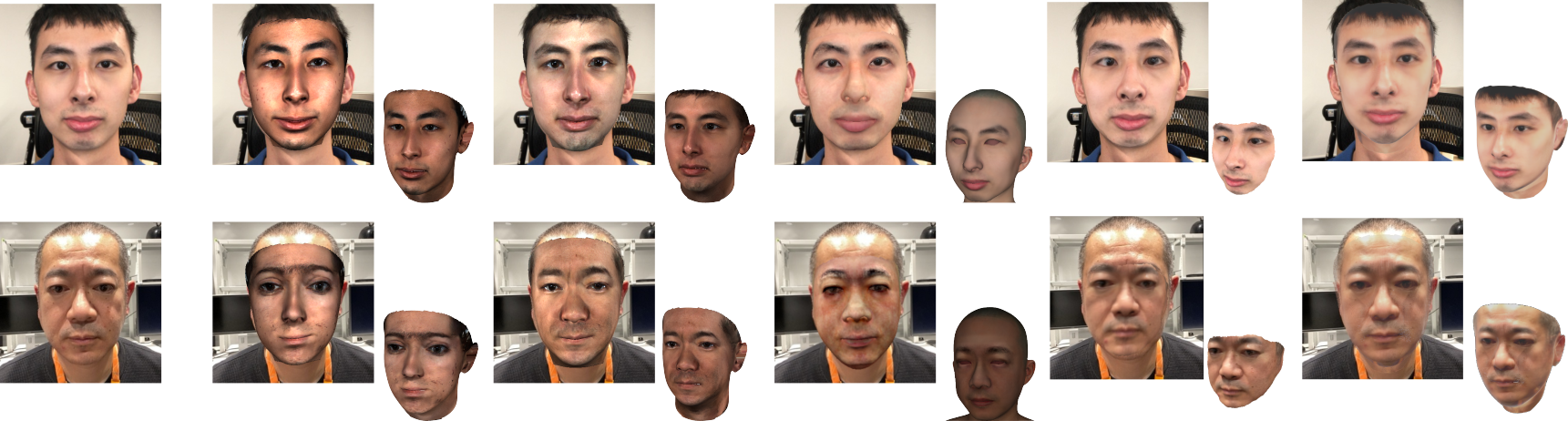}
\caption{
    Comparison with the existing face modeling methods.
    The top row shows the results with a public sequence ~\cite{hifi3dface2021tencentailab}, and the bottom row shows the results with our original sequence.
    GANFit \cite{gecer2019ganfit} and GANFit++ \cite{gecer2021fastganfit} are based on generative adversarial network (GAN) and DR.
    They use a single RGB input and estimate shape, texture, and lighting.
    hifi3dface \cite{hifi3dface2021tencentailab} with four view inputs minimizes a loss with 3DMM by DR and refines textures by GAN.
    COLMAP \cite{schoenberger2016sfm,schoenberger2016mvs} is a traditional SfM+MVS pipeline.
    We input 100 uniformly sampled RGB images and use additional face region masks for COLMAP.
    Its output mesh is textured by MVS-texturing \cite{Waechter2014Texturing}.
    Ours optimizes 3DMM by DR considering the normal and metallic-roughness textures with four views.
    Our results show a good similarity to subjects in the input images.
}
\label{fig:face_swap}
\EndFigureTwoCol

In this section, we demonstrate that Dressi can optimize a normal map to add fine details.
We optimize the normal map of two geometries to fit the rendered image using a ground truth normal map.
According to the norm constraint, we normalize the vector corresponding to each pixel every iteration.
The optimization results of the fish and boombox \cite{gltf} models are shown in \figurename~\ref{fig:normal_optimization}.
The normal map is initialized in a uniform light blue color
which represents the normal vector pointing directly to the viewer.
Then, we rotate the camera pose in the render for every frame to cover the geometry surface as much as possible.
In each frame, we calculate the RGB rendering loss  $L_{rbg}$ with Adam \cite{kingma2014method} optimizer.
The RGB rendering loss is the  $\ell_{2}$-norm of the difference between the rendered image using the ground truth normal map ${\mathbf N}_{GT}$ and that with the optimized normal map ${\mathbf N}_{\theta}$ parameterized by $\theta \in {\Theta}$.
In the upper part of \figurename~\ref{fig:normal_optimization}, the optimized normal map shows a similar pattern with the ground truth.
Considering the fin of the fish in \figurename~\ref{fig:normal_optimization} (a) and grooves of the boombox in \figurename~\ref{fig:normal_optimization} (b) as examples, the rendered images show the same detail with the ground truth, indicating the effectiveness of the normal map optimization.

To evaluate the dissimilarity between the optimized normal map ${\mathbf N}_{\theta}$ and the ground truth normal map ${\mathbf N}_{GT}$, we use the cosine dissimilarity value $L_{dissim}$ as a metric.
This is defined as:
\begin{equation}
    L_{dissim}(\theta)= \sum\limits_{p \in P({\mathbf N}_{GT})}  (1-cos({\mathbf N}_{\theta}(p),{\mathbf N}_{GT}(p)))/\|P({\mathbf N}_{GT})\|.
\end{equation}
where $P({\mathbf N}_{GT})$ is a set of pixel positions in a non-flat region and $p$ is a pixel position in the ground truth ${\mathbf N}_{GT}$.

Based on this metric, the bottom curve of \figurename~\ref{fig:normal_optimization} illustrates that the dissimilarity decreases in 1,200 iterations, and it finally converges to a small value.
The dissimilarity cannot converge to zero due to two reasons.
First, we fix the environment map as a uniform gray color during the optimization.
The optimized normal map will overfit this lighting condition.
Second, we rotate the camera pose uniformly.
There is no guarantee that every tiny surface is optimized because of the sparse views.
In spite of the limitations, the experimental results show that Dressi can optimize a pixel-wise normal map to enhance the fine details.

%% file: 9_appendix/5_faceswap.tex
\subsection{Human Face Modeling with 3D Morphable Model}
\label{subsec:face_swap}

To illustrate practical use of our approach in computer vision tasks, we apply Dressi to human face modeling using the 3D Morphable Model (3DMM) ~\cite{Egger20203DMF}.
In this experiment, we use a public selfie RGB-depth sequence ~\cite{hifi3dface2021tencentailab} and our original sequence captured in the same manner.
Their resolution is $480 \times 640$ and the intrinsic parameters are pre-calibrated.
In those sequences, a subject performs head rotation in front of a fixed camera.
Our method chooses four frames as input images by following \cite{hifi3dface2021tencentailab}.
Because a large difference in the rendering results is caused by view-direction, normal, and metallic-roughness components, it is good to use multiple views to estimate the view-independent values.
Our optimization process initializes albedo, normal, and metallic-roughness textures with flat values.
We initialize the camera poses of the four views and the scale of the 3DMM shape using the correspondences of detected 3D facial landmarks.
Then, we start the optimization of the camera poses, scale, PCA coefficients of 3DMM as blendshape weights, and the textures.
Our method uses SGD \cite{pmlr-v28-shamir13} to optimize the textures, and Adam \cite{kingma2014method} is used for the other parameters.
We use the fixed environment map as a uniform gray color and do not blur rendered images by HardSoftRas.
We minimize the loss $L$ as follows:
\begin{equation}
    L=w_lL_l + w_dL_d + w_{c}L_{c} + w_rL_r.
    \label{eqn:face_opt_loss}
\end{equation}
In this equation, $L_l$ is a 2D landmark loss with L1 norm, $L_d$ denotes a pixel-wise L1 depth loss with truncation, $L_c$ is a pixel-wise L2 RGB color loss, $L_r$ represents a regularization term for the PCA coefficients and those textures, and $w_l$, $w_d$, $w_c$, and $w_r$ are weights for the losses.
To render color images for $L_c$, we use PBS with albedo, normal, and metallic-roughness textures.
After the optimization, we apply simple image processing to the optimized textures to alleviate noises around the partially occluded areas for multiple views.
\figurename~\ref{fig:face_swap} shows the comparison with the existing methods.
Our method shows good geometry and texture reconstruction results with high visual similarity to the input image.
However, it is difficult for our optimization-based method to decompose materials correctly because of the fixed lighting and lack of data-driven priors.
For example, white highlights are often baked in albedo textures.
Lighting estimation \cite{10.1145/3130800.3130891}, learned features \cite{parkhi2015deep}, and material basis \cite{MobileSVBRDF:SIGA:2018, hifi3dface2021tencentailab} can improve material decomposition.